\documentstyle[12pt,twoside,epsf]{article}
\newcommand{\beq}{\begin{equation}}
\newcommand{\eeq}{\end{equation}}
\newcommand{\beqa}{\begin{eqnarray}}
\newcommand{\eeqa}{\end{eqnarray}}
\newcommand{\ymlsj}{Y^{M}_{LSJ}(\Omega )}
\newcommand{\oabs}[1]{|\vec{#1} \makebox[0.1em]{} |}
\newcommand{\oabsq}[1]{{|\vec{#1} \makebox[0.1em]{} |}^2}
\newcommand{\slsh}[1]{\, {\not {\! #1}}}
\newcommand{\slsha}[1]{\; {\not {\!\!\!\: #1}}}
\newcommand{\lams}{{\lambda }_{\sigma }^{0}}
\newcommand{\lamt}{{\lambda }_{\sigma }^{1}}
\newcommand{\epm}{\, \pm \hspace{- 0.69em} \bigcirc \,}
\newcommand{\emp}{\, \mp \hspace{- 0.69em} \bigcirc \,}
\newcommand{\xpm}{\, \pm \hspace{- 1.1em} \bigcirc \,}
\begin{document} 
\title{}
\author{}
\date{}
\maketitle
%\ \\[15mm]
%
%
% Ueberschrift und Autoren:
% =========================
%
{\LARGE \bf \raggedright Bethe--Salpeter--Approach to Relativistic \linebreak
Two--Fermion--Systems with a Separable \linebreak
Nonstatic Interaction} \\
       
{\large \raggedright F.\ Kleefeld and M.\ Dillig \\
{\small Institute for Theoretical Physics III, 
University of Erlangen--N\"urnberg, } \\
{\small Staudtstr.\ 7, D-91051 Erlangen, Germany 
\footnote{FAU-TP3-96/2} 
\footnote{Work supported by Kernforschungszentrum J\"ulich under contract No.\
ER--41154523} 
\footnote{Thanks to Prof.\ G.\ Rupp (Lisbon) for many fruitful discussions and
remarks}
\footnote{See also: Few-Body Systems 15 (1993) N24}} } \\

%
%
% Abstract:
% =========
%
\begin{quote}
\noindent {\bf Abstract.} To study the characteristic features of relativistic bound systems,
the Bethe-Salpeter equation (BSE) for two
equal mass spin $1/2$ particles (like the deuteron) is solved in the 
cm-frame for a covariant separable interaction kernel.
For that purpose the BSE is transformed to an eigenvalue problem which is
diagonalized numerically. The Bethe--Salpeter amplitudes (BSAs) are obtained 
straightforwardly from the resulting eigenvectors. Only positive
parity solutions of the eigenvalue problem are considered. To correlate the 
BSAs to standard quantum mechanical wavefunctions, the 
corresponding equal--time--wavefunctions (ETWs) are calculated. A 
decomposition of BSAs and ETWs in partial waves in angular 
momenta and parity is performed. 

As a first application
elastic electron--deuteron--scattering in the impulse approximation (IA) is considered.
The charge, magnetic and quadrupole formfactors $F_C(k^2)$, $F_M(k^2)$, $F_Q(k^2)$ and
tensor polarizations $\tilde{t}_{\,20}(k^2)$, $t_{\,20} \,(k^2,\theta_e=70^{\circ})$ are obtained
from three independent matrix elements of the deuteron current in 
the Breit--frame of elastic electron--deuteron--scattering. Additionally,
the formfactors $A(k^2)$ and $B(k^2)$ of the {\em Rosenbluth formula} are calculated.
\end{quote} 

%
%
% Haupttext:
% ==========
%
\section{Introduction}
The discussion of many body systems on a nuclear scale requires in
most cases a quantum mechanical and relativistic treatment. One approach,
most probably the correct description of such systems, is given by the
BSE. 

Unfortunately the practical solution of the BSE for most  
particle dynamics is highly non--trivial and extremely complex.
There are different ways out of this dilemma. One of them is to give up
covariance and to use static interaction kernels in a 
Schr\"odinger or Dirac like description of the many particle problem, excluding 
retardation effects in interaction. 

The way we follow in this note is to choose a covariant separable interaction kernel to satisfy
covariance and to simplify the mathematical treatment of the BSE. As we
are interested in bound systems we solve the homogenous BSE. Our test particle
is the deuteron as a weakly bound two--fermion--system with two
constituents, the proton and the neutron.

The choice of our phenomenological interaction kernel results in a  
manageable, though still complex mathematical problem. Goal of this paper is
the formulation and solution of the problem with a first simplistic application 
to elastic electron--deuteron--scattering. A more detailed
discussion of the formalism and applications to selected systems
will be presented elsewhere (e.g.\ the discussion of relativistic effects on bound state 
wavefuctions and formfactors or the investigation of antiparticle effects in 
few or many body systems at high energies).

%
% ==========================================================================
%
\section{Solution of the BSE in the rest frame of the deuteron}

Starting point is the homogenous BSE describing the deuteron
(mass $M_d$) as a bound system of a proton and a neutron (four---momenta $p_1$ and
$p_2$, mass $m$) in terms of the Jacobi four--momenta $P:=p_1+p_2$ and $q:=\frac{1}{2}\;(p_1-p_2)$:

\beq S_{F2}^{-1}(P,q)\:\Psi (P,q) = -\frac{i}{(2\pi )^4} \int \!\!d^4k\:K(P,q\,;k)\:\Psi (P,k) \label{lab0} \eeq

with

\beq S_{F2}(P,q) = {\left[ \,(\frac{1}{2} \slsha{P} + \slsh{q} - m + i\varepsilon )\otimes (\frac{1}{2} \slsha{P} - \slsh{q} - m + i\varepsilon )\,\right]}^{-1} \label{labx} \eeq

(to begin with we drop self--energy corrections to the nucleon mass $m$).

The most general parity conserving interaction kernel in terms of the various invariants of Dirac matrices is given as:

\beqa K(P\,,q\,;k) \;= \;\;\;\;\;g^2\cdot \, (& \!\!g_s \:\:f_s(P\,,q\,;k) & 1_4\otimes 1_4 \nonumber \\
                                    +& \!\!g_v \:\:f_v(P\,,q\,;k) & {\gamma }_{\mu }\otimes {\gamma }^{\mu } \nonumber \\
                                    +& \!\!g_t \:\:f_t(P\,,q\,;k) & {\sigma }_{\mu \nu}\otimes {\sigma }^{\mu \nu } \nonumber \\
                                    +& \!\!g_p \:\:f_p(P\,,q\,;k) & {\gamma }_5\otimes {\gamma }^5 \nonumber \\
                                    +& \!\!g_{pv} \,f_{pv}(P\,,q\,;k) & {\gamma }_5{\gamma }_{\mu }\otimes {\gamma }^5{\gamma }^{\mu }\:) 
\eeqa

For the practical calculation we use as interaction, respecting covariance, retardation effects and integrability of the Bethe--Salpeter--Equation (BSE), 
the following separable kernel ($q^2=(q^0)^2-\,\oabsq{q}$):

\beq K(q\,;k) \;= \;g^2\,( g_s \, v_s(q)\,v_s(k) \;{\Gamma }_s + g_v \, v_v(q)\,v_v(k) \;{\Gamma }_v + g_p \, v_p(q)\,v_p(k) \;{\Gamma }_p ) \label{lab1} \eeq

with

\beq  g_s=1 \qquad \makebox{\rm and } \qquad v_i(q) := \frac{{\Lambda }_i}{q^2 - {\Lambda }_i^2 + i\varepsilon } \makebox[2em]{ } (i = s,v,p) \label{lab2} \eeq

and with the spin--structures:

\beqa
 {\Gamma}_s & = & 1_4 \otimes 1_4 \nonumber \\
% & & \nonumber \\
 {\Gamma}_v & \stackrel{!}{=} & - \, {\gamma}^0 \otimes {\gamma}^0 + g_{\,0} \, \sum\limits_{i=1}^{3} {\gamma}^i \otimes {\gamma}^i \nonumber \\
% & & \nonumber \\
 {\Gamma}_p & = & {\gamma}^5 \otimes {\gamma}^5
\eeqa

The interaction parameters ${\Lambda }_s$, ${\Lambda }_v$, ${\Lambda }_p$ and the coupling constants $g_v, g_p, g_0$ are input parameters.
The coupling constant $g^2$ will be calculated from the eigenvalue condition.
%
% ==========================================================================
%
\subsection{Formulation of the eigenvalue problem}

Combination of the interaction kernel (\ref{lab1}) and the homogenous BSE (\ref{lab0})
yields:

\beq S_{F2}^{-1}(P,q)\:\Psi (P,q) = -\frac{i}{(2\pi )^4} \int \!\!d^4k\:\left( g^2\cdot \sum\limits_j g_j \, v_j(q)\,v_j(k) \;{\Gamma }_j\right) \:\Psi (P,k) \makebox[2em]{ } (j = s,v,p) \eeq

or after some transformations:

\beq \Psi (P,q) = -\frac{ig^2}{(2\pi )^4} S_{F2}(P,q)\:\sum\limits_j v_j(q)\,g_j \,{\Gamma }_j\:\int \!\!d^4k\:v_j(k) \;\Psi (P,k)  \label{lab3} \eeq

Multiplication by $v_i(q)$ and integration over $q$ leads to:

\beq \int \!\!d^4q\:v_i(q)\:\Psi (P,q) = -\frac{ig^2}{(2\pi )^4} \int \!\!d^4q\:S_{F2}(P,q)\:\sum\limits_j v_i(q)\,v_j(q)\,g_j \,{\Gamma }_j\:\int \!\!d^4k\:v_j(k) \;\Psi (P,k) \label{lab4} \eeq

\[ \makebox[25em]{ } (i,j = s,v,p) \]

For convenience we introduce the following quantities:

\beqa X_i    & := & \int \!\!d^4q\:v_i(q)\:\Psi (P,q) \\
             &    & \nonumber \\    
      A_{ij} & := & -\frac{i}{(2\pi )^4} \int \!\!d^4q\:S_{F2}(P,q)\:v_i(q)\,v_j(q) \label{lab5} \eeqa  
      
(Note that $X_i$ and $A_{ij}$ are functions of the total four--momentum $P$ !)

Using the definitions above the eigenvalue problem is obtained from (\ref{lab4}) as:

\beq \frac{1}{g^2} \:X_i = \sum\limits_j A_{ij}\,g_j \,{\Gamma }_j\:X_j \qquad \quad (i,j = s,v,p) \label{lab6} \eeq

or equivalently ($\lambda := g^{-2}$):

\beq 
\left( {\begin{array}{ccc} ( g_s A_{ss} {\Gamma}_s ) & (g_v A_{sv} {\Gamma}_v ) & ( g_p A_{sp} {\Gamma}_p ) \\ 
                           ( g_s A_{vs} {\Gamma}_s ) & (g_v A_{vv} {\Gamma}_v ) & ( g_p A_{vp} {\Gamma}_p ) \\ 
                           ( g_s A_{ps} {\Gamma}_s ) & (g_v A_{pv} {\Gamma}_v ) & ( g_p A_{pp} {\Gamma}_p ) \end{array} } \right)  
\left( {\begin{array}{c} X_s \\ X_v \\ X_p \end{array} } \right) 
\quad = \quad \lambda \; 
\left( {\begin{array}{c} X_s \\ X_v \\ X_p \end{array} } \right) 
\label{lab7} \eeq

The (unnormalized) BSAs $\Psi (P,q)$ are calculated from the eigenvectors $X_j$ and the corresponding
eigenvalues $g^{2}$ from eq.\ (\ref{lab3}) by:

\beq \Psi (P,q) = -\frac{ig^2}{(2\pi )^4} S_{F2}(P,q)\:\sum\limits_j v_j(q)\,g_j \,{\Gamma }_j\:X_j  \label{lab8} \eeq

Not all eigensolutions of the BSE are "physical" solutions; there are
e.g.\ solutions with negative norm. Unphysical solutions of the BSE are sometimes called
"Bethe--Salpeter ghosts" and are discussed e.g.\ in \cite{nak1}.

For our interaction kernel the normalization condition for the BSA reads:

\beq \pi i {\displaystyle \int } \! d^{\,4}q \:{\tilde{\Psi}}_{B'} \,(P,q) \, \frac{\partial}{\partial P_{\mu}} \, {\left[ S_{F2}(P,q) \right] }^{-1} {\biggr| }_{P^2=M^2_d} \,{\Psi}_B \,(P,q) \quad = \quad P^{\mu}\,{\delta}_{B'B} \label{lab9} \eeq

Here ${\tilde{\Psi}}$ is the adjoint BSA, $B$ and $B'$ are additional quantum numbers of the two--body bound states.

%
% ==========================================================================
%
\subsection{Calculation of ${A}_{ij}$ in the rest frame of the two--particle system}

For the numerical treatment of the eigenvalue problem (\ref{lab7})
the matrix ${A}_{ij}$ from eq.\ (\ref{lab5}) has to be calculated.
By the use of the definition of $v_i(q)$ in (\ref{lab2}) and the rationalized version of the
free two--fermion--propagator $S_{F2}(P,q)$ from (\ref{labx}): 

\beqa S_{F2}(P,q) & = & \frac{(\frac{1}{2} \slsha{P} + \slsh{q}  + m)\otimes (\frac{1}{2} \slsha{P} - \slsh{q} + m)}{((\frac{P}{2} + q)^2 - m^2 + i\varepsilon ) ((\frac{P}{2} - q)^2 - m^2 + i\varepsilon )} \label{lab10} \eeqa

the matrix ${A}_{ij}$ can be rewritten as:

\beq A_{ij} = -\frac{i}{(2\pi )^4} \int \!\!d^4q\:\frac{Z(P,q,m)}{N^{^{(ij)}}(P,q,m)} \cdot {\Lambda }_i{\Lambda }_j \label{lab11} \eeq

with

\beq Z(P,q,m) := {(\frac{1}{2} \slsha{P} + \slsh{q}  + m)\otimes (\frac{1}{2} \slsha{P} - \slsh{q} + m)} \label{lab12} \eeq

\beq N^{^{(ij)}}(P,q,m) := ((\frac{P}{2} + q)^2 - m^2 + i\varepsilon ) ((\frac{P}{2} - q)^2 - m^2 + i\varepsilon ) (q^2 - {\Lambda }_i^2 + i\varepsilon ) (q^2 - {\Lambda }_j^2 + i\varepsilon ) \label{lab13} \eeq

Multiplying out the numerator $Z(P,q,m)$ leads to:

\[ \begin{array}{ccrlc}
 A_{ij} \; \stackrel{!}{=} \; {\displaystyle \frac{{\Lambda}_i{\Lambda}_j}{(2\pi )^4}} & \biggl\{ \biggr. & (\frac{1}{4} P_{\mu}P_{\nu}\,I^{^{\;(ij)}} - I_{\mu \nu }^{^{\;(ij)}}) & ({\gamma }^{\mu}\otimes {\gamma }^{\nu}) & + \\
 & & & & \\
 & + & (\frac{m}{2}\,P_{\mu}\:I^{^{\;(ij)}}) & (({\gamma }^{\mu}\otimes 1_4) + (1_4\otimes {\gamma }^{\mu})) & + \\
 & & & & \\
 & + & (m^2\:I^{^{\;(ij)}}) & (1_4\otimes 1_4) & \biggl. \biggr\}
\end{array} \] \beq \label{lab14} \eeq

with the integrals:

\beqa I^{^{\;(ij)}}            & := & - i\:\int \!\!d^4q\:\frac{1}{N^{^{(ij)}}(P,q,m)} \nonumber \\
 & & \nonumber \\     
      I_{\mu }^{^{\;(ij)}}     & := & - i\:\int \!\!d^4q\:\frac{q_{\mu}}{N^{^{(ij)}}(P,q,m)} \quad \stackrel{!}{=} \quad 0 \nonumber \\
 & & \nonumber \\     
      I_{\mu \nu }^{^{\;(ij)}} & := & - i\:\int \!\!d^4q\:\frac{q_{\mu}q_{\nu}}{N^{^{(ij)}}(P,q,m)} \label{lab15} \eeqa

These integrations have been performed in the rest frame of the bound system; specifically for the deuteron  
$P^{\mu} = (M_d,\vec{0})$. The connection
between $M_d$ and $E_B$ ($E_B = $ binding energy of the deuteron) is:

\beq P^2 \quad = \quad (2\,m - E_B)^2 \quad = \quad M_d^2 \label{lab16} \eeq

%
% ==========================================================================
%
\subsection{Parity, angular momentum and the BSA}

The $16\times 16$--matrices $(g_jA_{ij}{\Gamma}_j)$\ $(i,j=s,v,p)$ in (\ref{lab7}) may be represented
by $4\times 4$--matrices with matrix elements being themselves $4\times 4$--matrices, consisting
of the $4\times 4$--unity--matrix $1_4$, the two particle spin operator $\sigma$ and its square ${\sigma}^2$.

The two particle spin operator $\sigma$ is defined as follows (${\sigma}^k$ with $k=1,2,3$ are the Pauli matrices in the 
z--representation):

\[ \begin{array}{ccccc} \sigma & = & \sum\limits_{k=1}^{3}  {\sigma }^k \otimes {\sigma }^k & = & {\left( \begin{array}{cccc} 1 & 0 & 0 & 0 \\
                                                                                                  0 & -1 & 2 & 0 \\
                                                                                                  0 & 2 & -1 & 0 \\
                                                                                                  0 & 0 & 0 & 1 \end{array} \right)} \end{array} \]

Its eigenvalues ${\lambda }_{\sigma }^{S}$ are:

\[ \begin{array}{cccl} \lams & = & -3 & {\rm (singlet)} \\
 & & \\                      
                       \lamt & = & 1  & {\rm (triplet)} \end{array} \]

The corresponding orthonormal eigenvectors ${\chi}_{m_S}^{S} $ are for the singlet ($S=0$) and for the triplet ($S=1$) spin states:

\[ {\chi}_0^0 = \frac{1}{\sqrt{2} } \left( \begin{array}{c} 0 \\ 1 \\ -1 \\ 0 \end{array} \right) , \makebox[3em]{ }
{\chi}_1^1 = \left( \begin{array}{c} 1 \\ 0 \\ 0 \\ 0 \end{array} \right) , \makebox[3em]{ }
{\chi}_0^1 = \frac{1}{\sqrt{2} } \left( \begin{array}{c} 0 \\ 1 \\ 1 \\ 0 \end{array} \right) , \makebox[3em]{ }
{\chi}_{-1}^1 = \left( \begin{array}{c} 0 \\ 0 \\ 0 \\ 1 \end{array} \right) \]

The subspace can be separated from the eigenvalue problem (\ref{lab7}) by the ansatz for the 
(16--component--)eigenvector $X_j$:

\beq X_j = X^{(S)}_j = \left( \begin{array}{c} X^{++(S)}_j \\ X^{+-(S)}_j \\ X^{-+(S)}_j \\ X^{--(S)}_j \end{array} \right) {\chi}_{m_S}^{S} \qquad (j=s,v,p) \eeq

The parity operator $\hat{{\cal P}}$ for the BSA is defined as:

\beq  \hat{{\cal P}} \quad := \quad ({\gamma}^0 \otimes {\gamma}^0) \;\;{\hat{{\cal P}}}^0 \label{parop} \eeq

with the so called "orbital parity operator" ${\hat{{\cal P}}}^0$\ :

\beq {\hat{{\cal P}}}^0 \; \Psi (P^0,\vec{P};q^0,\vec{q}) \quad = \quad \Psi (P^0, - \vec{P};q^0, - \vec{q}) \eeq

Finally, in our representation:

\beq {{\gamma }^0 \otimes {\gamma }^0} \quad = \quad {\left( \begin{array}{cccc} 1_4 & 0 & 0 & 0 \\
                                                                               0 & -1_4 & 0 & 0 \\
                                                                               0 & 0 & -1_4 & 0 \\
                                                                               0 & 0 & 0 & 1_4 \end{array} \right)} \label{gmm0} \eeq

It is easy to show the effect of the parity operator $\hat{{\cal P}}$ on the BSA of (\ref{lab8}) for $P^{\mu} = (M_d,\vec{0})$:

\beq \hat{{\cal P}}\;\Psi (P,q) \stackrel{!}{=} -\frac{ig^2}{(2\pi )^4} S_{F2}(P,q)\:\sum\limits_j v_j(q)\,g_j \,{\Gamma }_j\:({\gamma}^0 \otimes {\gamma}^0)\:X_j \qquad (j=s,v,p) \label{prit} \eeq

i.e. the parity of the BSA only depends on the operation of ${\gamma }^0 \otimes {\gamma }^0$ on $X_j$.
As the parity operator $\hat{{\cal P}}$ commutes with the interaction kernel $K(q\,;k)$ in (\ref{lab1}), there exist
eigensolutions of the BSA with positive and negative parity which can be selected (as
one can easily see from (\ref{prit}) and (\ref{gmm0})\ ) with the following $X_j$ ($j=s,v,p$):

\beq 
X_{+\,j}^{(S\,m_S)} :=  
 \left( {\begin{array}{c} X_j^{++(S)} \\ 0 \\ 0 \\ X_j^{--(S)}  
\end{array}} \right) {\chi}_{m_S}^S
\qquad ,\qquad 
X_{-\,j}^{(S\,m_S)} :=  
 \left( {\begin{array}{c} 0 \\ X_j^{+-(S)} \\ X_j^{-+(S)} \\ 0  
\end{array}} \right) {\chi}_{m_S}^S
\qquad \qquad (j=s,v,p)
\label{xxx} \eeq

For the deuteron only the states of positive parity (e.g. ${}^3S_1$ , ${}^3D_1$ , $\ldots$)
(spectroscopic notation: ${}^{2\,S+1}L_J$) are observed. For that reason we only
will consider eigensolutions of positive parity in the next few pages. For completeness we
mention that in case of negative parity states the eigenvalue problem (\ref{lab7}) can
be solved analytically.

For a decomposition of the BSA into angular momenta we
introduce the two fermion spherical harmonics $\ymlsj $ which couple
the spin angular momentum of a spin 1 particle (represented by the four--spinors ${\chi}_{m_S}^S$) 
to the orbital angular momentum (represented by the ordinary sperical harmonics $Y_{Lm_L}(\Omega )$):

%\newpage

\beq \ymlsj = \sum_{m_L,m_S} <LSm_Lm_S|JM> \, Y_{Lm_L}(\Omega )\, \chi_{m_S}^{S} \eeq

with the orthonormality relation:

\beq \int \!\!d\Omega\, Y_{L'\!S'\!J'}^{M'\ \dag}(\Omega )\,\ymlsj  = {\delta }_{M'\!M}\,{\delta }_{L'\!L}\,{\delta }_{S'\!S}\,{\delta }_{J'\!J} \eeq

(The Clebsch--Gordan--coefficients follow the convention
of Condon \& Shortley \cite{con1}). 

The application of the projections:

\beqa {\chi }_M^J & = & \sqrt{4\pi } \, Y_{0JJ}^M(\Omega ) \makebox[0.5em]{ } (J=0,\:M=0\: {\rm or}\: J=1,\:M=0,\pm 1) \label{zwsp1} \\
 & & \nonumber \\
(\frac{\vec{\sigma } \cdot \vec{q} }{\oabs{q} } \otimes 1_2) \, {\chi }_0^0 & = & - \sqrt{4\pi } \, Y_{110}^0(\Omega ) \\
 & & \nonumber \\
(\frac{\vec{\sigma } \cdot \vec{q} }{\oabs{q} } \otimes 1_2) \, {\chi }_M^1 & = & \sqrt{\frac{4\pi }{3}} \, (Y_{101}^M(\Omega ) - \sqrt{2} \, Y_{111}^M(\Omega )) \makebox[2em]{ } (M=0,\pm 1) \\
 & & \nonumber \\
(1_2 \otimes \frac{\vec{\sigma } \cdot \vec{q} }{\oabs{q} } ) \, {\chi }_0^0 & = & \sqrt{4\pi } \, Y_{110}^0(\Omega ) \\
 & & \nonumber \\
(1_2 \otimes \frac{\vec{\sigma } \cdot \vec{q} }{\oabs{q} } ) \, {\chi }_M^1 & = & \sqrt{\frac{4\pi }{3}} \, (-Y_{101}^M(\Omega ) - \sqrt{2} \, Y_{111}^M(\Omega )) \makebox[2em]{ } (M=0,\pm 1) \\
 & & \nonumber \\
(\frac{\vec{\sigma } \cdot \vec{q} }{\oabs{q} } \otimes \frac{\vec{\sigma } \cdot \vec{q} }{\oabs{q} } ) \, {\chi }_0^0 & = & - \sqrt{4\pi } \, Y_{000}^0(\Omega ) \\
 & & \nonumber \\
(\frac{\vec{\sigma } \cdot \vec{q} }{\oabs{q} } \otimes \frac{\vec{\sigma } \cdot \vec{q} }{\oabs{q} } ) \, {\chi }_M^1 & = & \frac{\sqrt{4\pi }}{3} \, (Y_{011}^M(\Omega ) + \sqrt{8} \, Y_{211}^M(\Omega )) \makebox[2em]{ } (M=0,\pm 1) \label{zwsp2} \eeqa

to the positive parity BSA combining (\ref{lab8}) and (\ref{xxx}):

\beq {\Psi}^{(JM)} (P,q) = -\frac{i(g^{(J)}(P))^2}{(2\pi )^4} S_{F2}(P,q)\:\sum\limits_j v_j(q)\,g_j \,{\Gamma }_j\:X_{+\,j}^{(JM)}(P) \qquad (j=s,v,p) \label{ppp} \eeq

leads to the following decomposition of the positive parity BSA in partial waves:

\beqa
 {\Psi}^{\,(00)}\,(M_d\, , \vec{0} ; q) & = & \!\left( {\begin{array}{c} 
 {\Psi}_{\displaystyle {\,}^1S_0^{+}}\,(M_d\, ; q_0 , \oabs{q}) \;Y_{000}^0(\Omega ) \\ \\
 {\Psi}_{\displaystyle {\,}^3P_0^{+}}\,(M_d\, ; q_0 , \oabs{q}) \;Y_{110}^0(\Omega ) \\ \\
 {\Psi}_{\displaystyle {\,}^3P_0^{-}}\,(M_d\, ; q_0 , \oabs{q}) \;Y_{110}^0(\Omega ) \\ \\
 {\Psi}_{\displaystyle {\,}^1S_0^{-}}\,(M_d\, ; q_0 , \oabs{q}) \;Y_{000}^0(\Omega ) 
 \end{array} } \right) \\
 & & \nonumber \\
 & & \nonumber \\
 {\Psi}^{\,(1\,M)}\,(M_d\, , \vec{0} ; q) & = & \!\left( {\begin{array}{ccc} 
 {\Psi}_{\displaystyle {\,}^3S_1^{+}}\,(M_d\, ; q_0 , \oabs{q}) \;Y_{011}^M(\Omega ) & + & {\Psi}_{\displaystyle {\,}^3D_1^{+}}\,(M_d\, ; q_0 , \oabs{q}) \;Y_{211}^M(\Omega ) \\ & & \\
 {\Psi}_{\displaystyle {\,}^3P_1^{+}}\,(M_d\, ; q_0 , \oabs{q}) \;Y_{111}^M(\Omega ) & + & {\Psi}_{\displaystyle {\,}^1P_1^{+}}\,(M_d\, ; q_0 , \oabs{q}) \;Y_{101}^M(\Omega ) \\ & & \\
 {\Psi}_{\displaystyle {\,}^3P_1^{-}}\,(M_d\, ; q_0 , \oabs{q}) \;Y_{111}^M(\Omega ) & + & {\Psi}_{\displaystyle {\,}^1P_1^{-}}\,(M_d\, ; q_0 , \oabs{q}) \;Y_{101}^M(\Omega ) \\ & & \\
 {\Psi}_{\displaystyle {\,}^3S_1^{-}}\,(M_d\, ; q_0 , \oabs{q}) \;Y_{011}^M(\Omega ) & + & {\Psi}_{\displaystyle {\,}^3D_1^{-}}\,(M_d\, ; q_0 , \oabs{q}) \;Y_{211}^M(\Omega ) 
 \end{array} } \right) 
 \nonumber \\
 & & \nonumber \\
 & & \nonumber \\
 & & 
\eeqa

in the rest frame of the deuteron.

%
% ==========================================================================
%
\section{The equal-time-wavefunction $\Phi \,(P, \vec{q})$ (ETW)}

To correlate the determined BSAs to conventional bound state wavefunctions,
the corresponding ETWs $\Phi \,(P, \vec{q})$ are computed by:

\beq \Phi \,(P, \vec{q}) \quad := \quad \int \! dq^0 \:\:\Psi (P,q) \: e^{\displaystyle i\,q^0x^0} { \Biggr| }_{\displaystyle x^0 = 0} \quad \stackrel{!}{=} \quad \int \! dq^0 \:\:\Psi (P,q) \eeq

With the definition of:

\beq I_{_{(i)}} (P,\vec{q}) \quad := \quad \int \! dq^0 \:\:S_{F2}\,(P,q)\,v_i(q) \qquad (i=s,v,p) \label{iidf1} \eeq

the BSA (\ref{ppp}) changes in the rest frame to:

\beq {\Phi}^{(JM)} (M_d\, ,\vec{0};\vec{q}) = -\frac{i(g^{(J)}(M_d\, ,\vec{0}))^2}{(2\pi )^4} \:\sum\limits_j I_{_{(j)}} (M_d\, ,\vec{0};\vec{q})\,g_j \,{\Gamma }_j\:X_{+\,j}^{(JM)}(M_d\, ,\vec{0}) \!\!\quad (j=s,v,p) \label{ppp1} \eeq

The decomposition in partial waves gives for the positive parity ETW in the rest frame:

\beqa
 {\Phi}^{\,(00)}\,(M_d\, ,\vec{0}; \vec{q}) & = & \left( {\begin{array}{c} 
 {\Phi}_{\displaystyle {\,}^1S_0^{+}}\,(M_d\, ; \oabs{q}) \;Y_{000}^0(\Omega ) \\ \\
 {\Phi}_{\displaystyle {\,}^3P_0^{+}}\,(M_d\, ; \oabs{q}) \;Y_{110}^0(\Omega ) \\ \\
 {\Phi}_{\displaystyle {\,}^3P_0^{-}}\,(M_d\, ; \oabs{q}) \;Y_{110}^0(\Omega ) \\ \\
 {\Phi}_{\displaystyle {\,}^1S_0^{-}}\,(M_d\, ; \oabs{q}) \;Y_{000}^0(\Omega ) 
 \end{array} } \right) 
 \qquad \label{par1} \\
 & & \nonumber \\
 & & \nonumber \\
 {\Phi}^{\,(1\,M)}\,(M_d\, ,\vec{0} ; \vec{q}) & = & \left( {\begin{array}{ccc} 
 {\Phi}_{\displaystyle {\,}^3S_1^{+}}\,(M_d\, ; \oabs{q}) \;Y_{011}^M(\Omega ) & + & {\Phi}_{\displaystyle {\,}^3D_1^{+}}\,(M_d\, ; \oabs{q}) \;Y_{211}^M(\Omega ) \\ & & \\
 {\Phi}_{\displaystyle {\,}^3P_1^{+}}\,(M_d\, ; \oabs{q}) \;Y_{111}^M(\Omega ) & + & {\Phi}_{\displaystyle {\,}^1P_1^{+}}\,(M_d\, ; \oabs{q}) \;Y_{101}^M(\Omega ) \\ & & \\
 {\Phi}_{\displaystyle {\,}^3P_1^{-}}\,(M_d\, ; \oabs{q}) \;Y_{111}^M(\Omega ) & + & {\Phi}_{\displaystyle {\,}^1P_1^{-}}\,(M_d\, ; \oabs{q}) \;Y_{101}^M(\Omega ) \\ & & \\
 {\Phi}_{\displaystyle {\,}^3S_1^{-}}\,(M_d\, ; \oabs{q}) \;Y_{011}^M(\Omega ) & + & {\Phi}_{\displaystyle {\,}^3D_1^{-}}\,(M_d\, ; \oabs{q}) \;Y_{211}^M(\Omega ) 
 \end{array} } \right) 
 \qquad \label{par2} 
\eeqa

The free two--fermion--propagator can be expanded with the help
of the two fermion energy projection operators 
${\Lambda}^{\pm \epm}(\vec{q}) := {\Lambda}^{\pm}(\vec{q}) \,\otimes {\Lambda}^{ \epm }(- \,\vec{q})$ :

\beqa 
 \lefteqn{S_{F2}(M_d\, ,\vec{0}\,;q) \quad = } \nonumber \\
 & & \nonumber \\
 & & \nonumber \\
 & = & \left( \sum\limits_{\pm \epm } \frac{{\Lambda}^{\pm \epm}(\vec{q})}{((E_N + q^0) \pm (- \, \omega (\vec{q}) + i\varepsilon ))\,((E_N - q^0) \xpm (- \, \omega (\vec{q}) + i\varepsilon ))} \right)
 \, \left( {\gamma}^0 \otimes {\gamma}^0 \right) \quad \\
 & & \nonumber \\
 & & \nonumber 
\eeqa

\[ (E_N := \frac{M_d}{2} \quad , \quad \omega (\vec{q}) := \sqrt{\oabsq{q} + m^2} \quad , \quad {\omega}_i (\vec{q}) := \sqrt{\oabsq{q} + {\Lambda}_i^2} \quad ) \label{zflll} \]

Upon performing the integration in (\ref{iidf1}), we find
in the rest frame of the bound state:

\beqa
 I_{_{(i)}} (M_d\, ,\vec{0}\,;\vec{q}) & = \quad {\displaystyle \frac{2 \, \pi }{i}} \quad \cdot & 
 \Bigl( \Bigr. \quad w_{_{(i)}}^{++}(E_N ; \vec{q}) \;{\Lambda}^{++}(\vec{q}) \quad + \nonumber \\
 & & \nonumber \\
 & & + \quad w_{_{(i)}}^{+-}(E_N ; \vec{q}) \;{\Lambda}^{+-}(\vec{q}) \quad + \nonumber \\
 & & \nonumber \\
 & & + \quad w_{_{(i)}}^{-+}(E_N ; \vec{q}) \;{\Lambda}^{-+}(\vec{q}) \quad + \nonumber \\
 & & \nonumber \\
 & & + \quad w_{_{(i)}}^{--}(E_N ; \vec{q}) \;{\Lambda}^{--}(\vec{q}) \Bigl. \Bigr) \:\left( {\gamma}^0 \otimes {\gamma}^0 \right) 
\eeqa

with

\beqa 
 w_{_{(i)}}^{++}(E_N ; \vec{q}) & := & \frac{{\Lambda}_i}{2\,(E_N - \omega (\vec{q}))\,((E_N - \omega (\vec{q}))^2 - {\omega}_i^2 (\vec{q}))} \quad + \nonumber \\
 & & \nonumber \\
 & & \nonumber \\
 & + & \frac{{\Lambda}_i}{((E_N - \omega (\vec{q}))^2 - {\omega}_i^2 (\vec{q}))} \: \cdot \: \frac{1}{2\,{\omega}_i (\vec{q})} \nonumber \\
 & & \nonumber \\
 & & \nonumber \\
 w_{_{(i)}}^{+-}(E_N ; \vec{q}) & := & \frac{{\Lambda}_i}{(E_N^2 - (\omega (\vec{q}) + {\omega}_i (\vec{q}))^2)} \: \cdot \: \frac{1}{2\,{\omega}_i (\vec{q})} \nonumber \\
 & & \nonumber \\
 & & \nonumber \\
 w_{_{(i)}}^{-+}(E_N ; \vec{q}) & := & \frac{{\Lambda}_i}{(E_N^2 - (\omega (\vec{q}) + {\omega}_i (\vec{q}))^2)} \: \cdot \: \frac{1}{2\,{\omega}_i (\vec{q})} \nonumber \\
 & & \nonumber \\
 & & \nonumber \\
 w_{_{(i)}}^{--}(E_N ; \vec{q}) & := & \frac{- \,{\Lambda}_i}{2\,(E_N + \omega (\vec{q}))\,((E_N + \omega (\vec{q}))^2 - {\omega}_i^2 (\vec{q}))} \quad + \nonumber \\
 & & \nonumber \\
 & & \nonumber \\
 & + & \frac{{\Lambda}_i}{((E_N + \omega (\vec{q}))^2 - {\omega}_i^2 (\vec{q}))} \: \cdot \: \frac{1}{2\,{\omega}_i (\vec{q})} 
\eeqa

\begin{figure}[p]
\epsfxsize=  17.0cm
\epsfysize=  10.0cm
\centerline{\epsffile{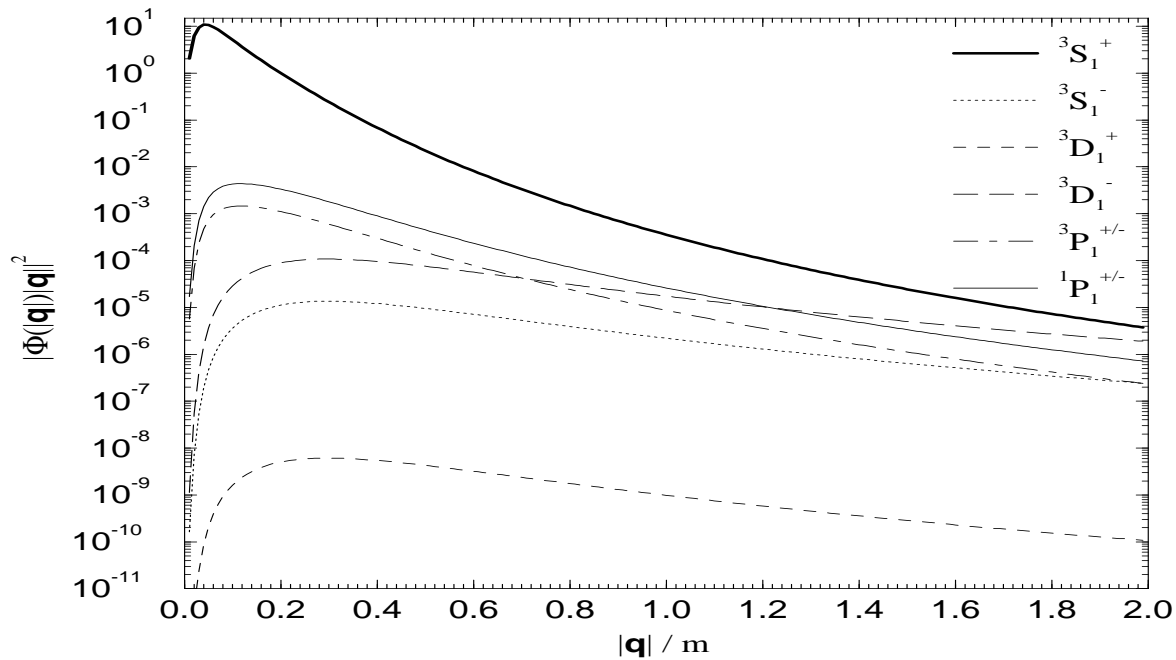}}
\caption{Partial wave decomposition of the $J^{\pi}=1^{+}$ Bethe--Salpeter amplitude for 
${\Lambda}_s={\Lambda}_v={\Lambda}_p=0.34 \:$ , $\: g_v=g_0=g_p=0 \:$ , $E_B=0.002$ (the
cut--off masses ${\Lambda}_i$ and the binding energy $E_B$ are given in units of the nucleon
mass) \label{fff1}} 
\end{figure}

\begin{figure}[p]
\epsfxsize=  17.0cm
\epsfysize=  10.0cm
\centerline{\epsffile{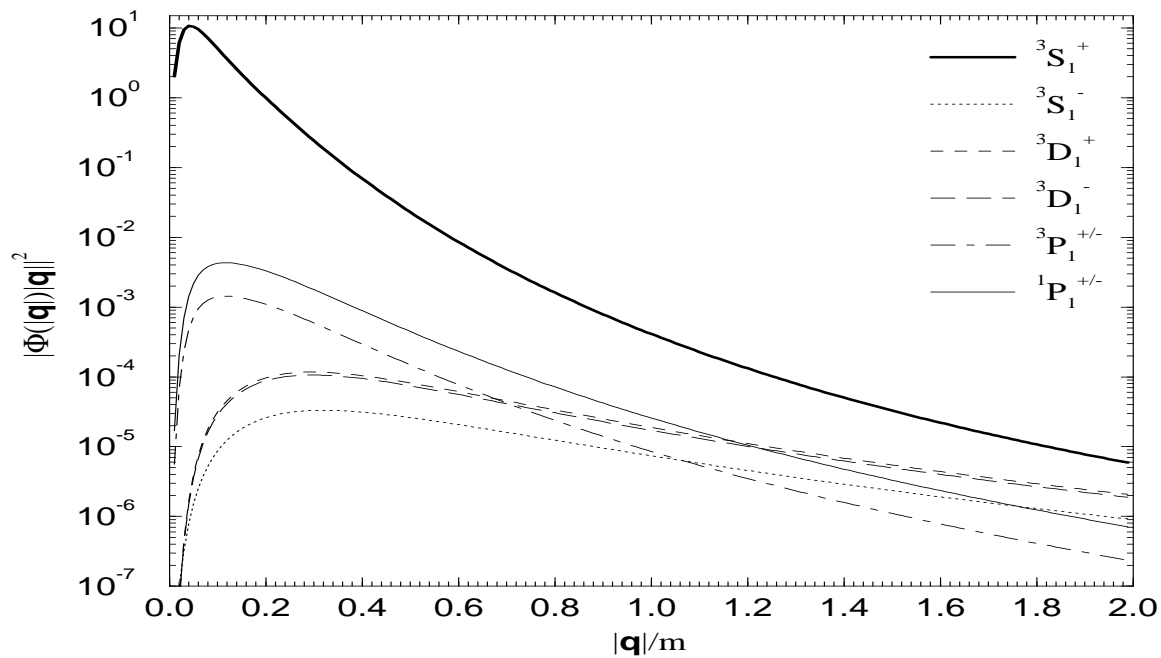}}
\caption{As Fig.\ 1, however for the parameters: ${\Lambda}_s={\Lambda}_v={\Lambda}_p=0.34 \:$ , $\: g_v=1.2 \:$ , $\: g_0=1. \:$ , $\: g_p=-0.99 \:$ , $E_B=0.002$ \label{fff2}} 
\end{figure}

Explicit expressions for the partial waves of equation (\ref{par1})
and (\ref{par2}) are given in appendix \ref{app1}.
Partial waves of the ETW in the rest frame (in arbitary units) are shown in Fig. \ref{fff1} (scalar interaction) and
Fig. \ref{fff2} (full interaction) for one selected eigensolution
of the BSE. In the presented examples one can see the strong dependence of the ${}^3D^+_1$ wave
and the antiparticle content of the BSA represented e.g.\ by the ${}^3S^-_1$ wave on
the choice of the coupling constants in our model.

%
% ==========================================================================
%
\section{Elastic e-d-scattering in the IA and deuteron form factors}

As a first application, we consider elastic electron--deuteron--scattering in impulse approximation (IA),
following a similar route as in M.J.\ Zuilhof and J.A.\ Tjon 1980
\cite{zui1}. The connection between the BSA and the deuteron current matrix
elements in the IA is given as \cite{mic1}:

\beqa
 \lefteqn{ <\!P_f,M_f|j_d^{\mu}|P_i,M_i\!> \quad =} \nonumber \\
 & & \nonumber \\
 & \stackrel{!}{=} & 2\pi i\,\frac{e}{M_d} \, {\displaystyle \int \!\! d^{\,4}q }\;{\tilde{\Psi}}_{M_f} (P_f, \, q + \frac{1}{2}\,k)\;\left( {\Gamma}^{\mu}(k^2)\;\otimes \;
 (\frac{1}{2}\,{\slsha{P}}_i - \slsh{q} \, - \,m ) \right) \:{\Psi}_{M_i} (P_i,q) \quad \quad 
\eeqa

Thereby, $M_i$ and $M_f$ is the polarization of the deuteron before and after scattering, respectively, $k$ is the four--momentum transfer by the photon.
The vertex function ${\Gamma}^{\mu}(k^2)$ is evaluated with the isoscalar (dipole) formfactors $F_1^S(k^2)$ and $F_2^S(k^2)$ of the nucleon:

\beq {\Gamma}^{\mu}(k^2) \quad = \quad {\gamma}^{\mu} \: F_1^S(k^2) \: + \:\: \frac{i}{2m} \,{\sigma}^{\mu \nu}\,k_{\nu} \:F_2^S(k^2) \eeq

For the nucleonic formfactors $F_1^S(k^2)$ and $F_2^S(k^2)$ we use fits of Iachello et al. \cite{iac1}.

From the current matrix elements obtained we can calculate the observables of the
deuteron (e.g.\ formfactors). First we introduce the following
covariant and contravariant spherical unit vectors $\vec{\varepsilon}_M$ and
$\vec{\varepsilon}^{\,M}$ ($M=+1,0,-1$) expressed in terms of the 
cartesian ones ($\vec{e}_x$,$\vec{e}_y$,$\vec{e}_z$):

\[  \vec{\varepsilon}_{\pm 1} := \mp \frac{1}{\sqrt{2}} (\vec{e}_x \pm i \vec{e}_y) =: {(\vec{\varepsilon}^{\,\pm 1})}^\ast \quad , 
\quad \vec{\varepsilon}_{0} := \vec{e}_z =: {(\vec{\varepsilon}^{\,0})}^\ast \]

with the properties ($M,M'=+1,0,-1$):

\[ \vec{\varepsilon}_{M} = {(-1)}^M  \vec{\varepsilon}^{\,\,-M} \quad 
\mbox{and} \quad  \vec{\varepsilon}^{\,M} \cdot \vec{\varepsilon}_{M'} = {\delta}^{M}_{M'} \] 

Contravariant and covariant components of spherical vectors are defined by $\vec{A} = A^{\,M} \vec{\varepsilon}_{M} = A_M \vec{\varepsilon}^{\,M} $.
Now we can define the polarization vectors for a massive spin 1 particle like the deuteron:

\[ {\varepsilon}^{\,\mu \,M} (P) = ({\varepsilon}^{\,0 \,M} (P) \, , \, \vec{\varepsilon}^{\,M} (P) ) := 
\left( \frac{\vec{P} \cdot \vec{\varepsilon}^{\,M}}{M_d} \, , \, \vec{\varepsilon}^{\,M} 
+ \frac{\vec{P} \cdot \vec{\varepsilon}^{\,M}}{M_d \,(P^0 + M_d)} \,\vec{P} \right) \]

where $\mu$ is a Lorentz index with $\mu =0,1,2,3$. The properties of the polarization
vectors are well known:

\beqa {\varepsilon}^{\,M} (P) \cdot {\varepsilon}_{M'} (P) \quad = \quad {\varepsilon}^{\,\mu \,M} (P) \, {\varepsilon}_{\mu\,M'} (P) & = & - \, {\delta}^{M}_{M'} \nonumber \\
 & & \nonumber \\
 \sum_M {({\varepsilon}_{\mu \,M} (P))}^{\ast} \, {\varepsilon}_{\nu \,M} (P) \quad = \quad \,\,\,
 {\varepsilon}_{\mu}^{\,\,\,M} (P) \, {\varepsilon}_{\nu \,M} (P) & = & - \, g_{\mu \nu} + \frac{P_{\mu}P_{\nu}}{M_d^{\,2}} \nonumber \\
 & & \nonumber \\
 P^{\,\mu} {\varepsilon}_{\mu \,M} (P) & = & 0 \nonumber 
 \eeqa 
 
In terms of the momentum transfer $k = P_f - P_i$ and the polarization vectors the
current matrix elements of a massive spin 1 particle can be expressed in the following
covariant way \cite{gla1} \cite{zui1} \cite{rup1} \cite{hum1}:

\beq <\!P_f,M_f|j_{\mu}^d|P_i,M_i\!> \quad = \quad - \, \frac{e}{2\,M_d} \: {\varepsilon}_{\rho \,M_f}^{\ast} (P_f) \:J_{\mu}^{\rho \sigma} \: {\varepsilon}_{\sigma M_i} (P_i) \label{strm1} \eeq

with the current tensor:

\beq J_{\mu}^{\rho \sigma} \quad = \quad (P_{f,\mu} + P_{i,\mu}) \left[ g^{\rho \sigma} F_1(k^2) \: - \: \frac{k^{\rho} k^{\sigma}}{2\,M_d^2} \,F_2\,(k^2) \right] + i \,I_{\mu \nu}^{\rho \sigma} k^{\nu} \,G_1(k^2) \eeq

Here $F_1(k^2)$, $F_2(k^2)$, $G_1(k^2)$ are formfactors and $I_{\mu \nu}^{\rho \sigma} = i\,(g^{\,\rho}_{\mu} g^{\,\sigma}_{\nu} - g^{\,\rho}_{\nu} g^{\,\sigma}_{\mu})$ are the generators of the Lorentz group.
Going to the Breit--frame ($k^0=0$) and choosing $\vec{k}$ in the $z$--direction, i.e.\ $k^{\,\mu}=(0,0,0,k_z)$,
evluation of equation (\ref{strm1}) with respect to elastic scattering leads to the following relation
between the formfactors and the current matrix elements ($\eta := - k^2 / (2\,M_d)^2$):

\beqa
 <\!P_f,M_f|j_d^0|P_i,M_i\!> & = & e \:\sqrt{1 + \eta} \: \biggl\{ F_1 \,{\delta}^{\,M_f}_{M_i} + \nonumber \\
 & & \nonumber \\
 & + & 2 \, \eta \: \left[ F_1 + (1 + \eta ) \, F_2 - G_1 \right] \,{\delta}^{\,M_f}_{\,0}\,{\delta}^{\,0}_{M_i} \biggr\} \\
 & & \nonumber \\
 <\!P_f,M_f|j_d^1|P_i,M_i\!> & = & e \,\frac{k_z}{2\,M_d} \sqrt{\frac{1 + \eta}{2}} \:G_1\:({\delta}^{\,M_f}_{M_i+1} - \,{\delta}^{\,M_f}_{M_i-1}) \\
 & & \nonumber \\
 <\!P_f,M_f|j_d^2|P_i,M_i\!> & = & - \,ie \,\frac{k_z}{2\,M_d} \sqrt{\frac{1 + \eta}{2}} \:G_1\:({\delta}^{\,M_f}_{M_i+1} + {\delta}^{\,M_f}_{M_i-1}) \\
 & & \nonumber \\
 <\!P_f,M_f|j_d^3|P_i,M_i\!> & = & 0 \label{cont1}
\eeqa

Equation (\ref{cont1}) is the continuity equation for the
deuteron current in our frame of reference.

The charge, magnetic and quadrupole formfactors $F_C(k^2)$, $F_M(k^2)$, $F_Q(k^2)$ are
related to the formfactors $F_1(k^2)$, $F_2(k^2)$, $G_1(k^2)$ by (see e.g.\ \cite{gou1}):

\beqa
 F_C(k^2) & = & F_1(k^2) + \frac{2}{3} \,\eta \, \left[ F_1(k^2) + (1 + \eta ) \, F_2\,(k^2) - G_1(k^2) \right] \\
 & & \nonumber \\
 F_M(k^2) & = & G_1(k^2) \\
 & & \nonumber \\
 F_Q(k^2) & = & F_1(k^2) + (1 + \eta ) \, F_2\,(k^2) - G_1(k^2) 
\eeqa

At $k^2=0$ they have the following values \cite{gar1}:

\beq e\,F_C(0) \, = \, 1\,e \qquad , \qquad \frac{e}{2\,M_d} \, F_M(0) \, = \, {\mu}_d  \qquad , \qquad \frac{e}{M_d^2} \, F_Q(0) \, = \, Q_d \eeq

$1\,e$, ${\mu}_d$ and $Q_d$ are the deuteron's charge, magnetic and quadrupole moment 
(we note that in \cite{gla1} the quadrupole moment is defined by $Q_d=\frac{e}{M_d^2} \, F_2(0)$).
Hence in the Breit-frame $F_C(k^2)$, $F_M(k^2)$, $F_Q(k^2)$ are obtained
from three independent matrix elements of the deuteron current, e.g.:

\beqa
             e \:\sqrt{1 + \eta} \: F_C & = & \frac{1}{3} <\!P_f,0|j_d^0|P_i,0\!> \: + \: \frac{2}{3} <\!P_f,1|j_d^0|P_i,1\!> \\
 & & \nonumber \\
 e \,\frac{k_z}{2\,M_d} \sqrt{\frac{1 + \eta}{2}} \:F_M & = & \quad <\!P_f,1|j_d^1|P_i,0\!> \\
 & & \nonumber \\
 2 \, \eta \,e \:\sqrt{1 + \eta} \: F_Q & = & \quad <\!P_f,0|j_d^0|P_i,0\!> \: - \: \quad <\!P_f,1|j_d^0|P_i,1\!> 
\eeqa

The formfactors $A(k^2)$, $B(k^2)$ of the {\em Rosenbluth formula} are calculated easily by (see e.g.\ \cite{gar1}):

\beq A(k^2) = F_C^2 + \frac{8}{9} \,{\eta}^2\,F_Q^2 + \frac{2}{3} \,\eta \, F_M^2 \quad , \quad 
 B(k^2) = \frac{4}{3} \, \eta \,(\eta + 1)\:F_M^2 \label{aabb} \eeq

with the appropriately boosted BSAs and deuteron currents from the deuteron rest frame to the Breit--frame. 
Knowing $F_C(k^2)$, $F_M(k^2)$, $F_Q(k^2)$ it is straightforward to calculate e.g.
the simplified tensor polarization $\tilde{t}_{\,20}$ discussed in \cite{gar1}:

\beq \tilde{t}_{\,20} = - \, \sqrt{2} \, \frac{x (x+2)}{1+2x^2} \quad \mbox{with} \quad x= \frac{2 \eta F_Q}{3 F_C} \label{tt20} \eeq

and the complete (observable) tensor polarization $t_{\,20}$ defined by:

\beq t_{\,20} = - \, \sqrt{2} \,\, \frac{x (x+2) + y/2}{1+2\,(x^2+y)} \quad \mbox{with} \quad y= \frac{2 \eta}{3} \,(\,\frac{1}{2} + (1 + \eta ) \, \tan^2{} \frac{{\theta}_e}{2}  ) \,\, {\left( \frac{F_M}{F_C} \right) }^2 \label{ctt20} \eeq

\begin{figure}[p]
\epsfxsize=  17.0cm
\epsfysize=  10.0cm
\centerline{\epsffile{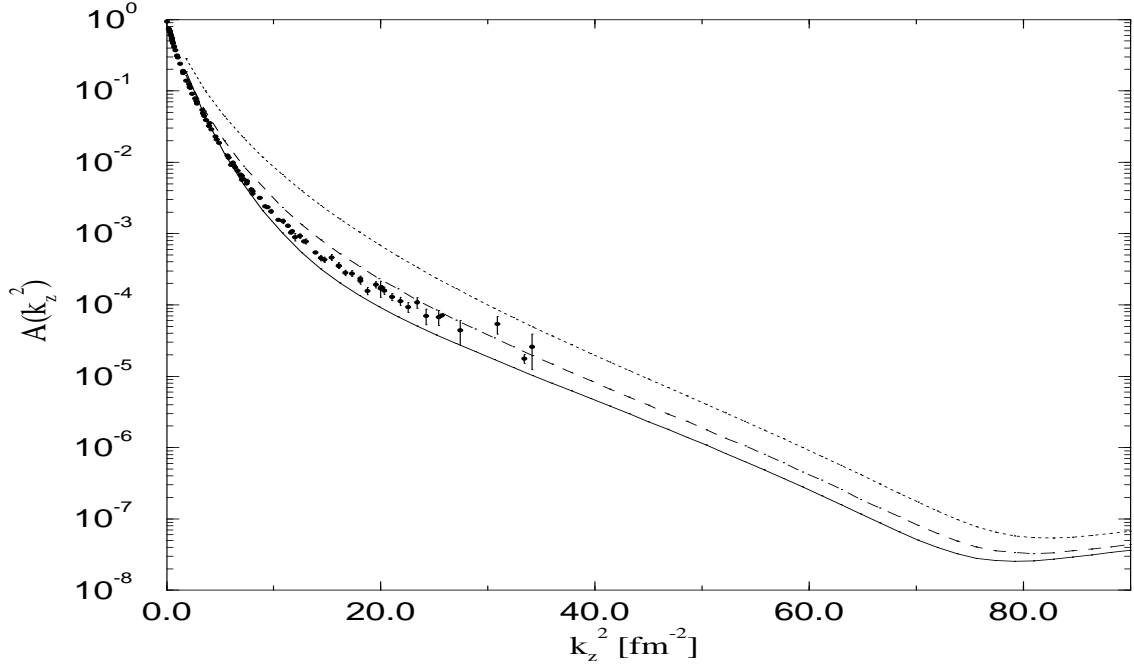}}
\caption{Dependence of the formfactor $A(k_z^2)$ from equ.\ (64)
on the momentum transfer $k^2=-k_z^2$ (The sets of model parameters
compared are summarized in Tab.\ 1; for experimental data 
see appendix B) \label{fffa}} 
\end{figure}

\begin{figure}[p]
\epsfxsize=  17.0cm
\epsfysize=  10.0cm
\centerline{\epsffile{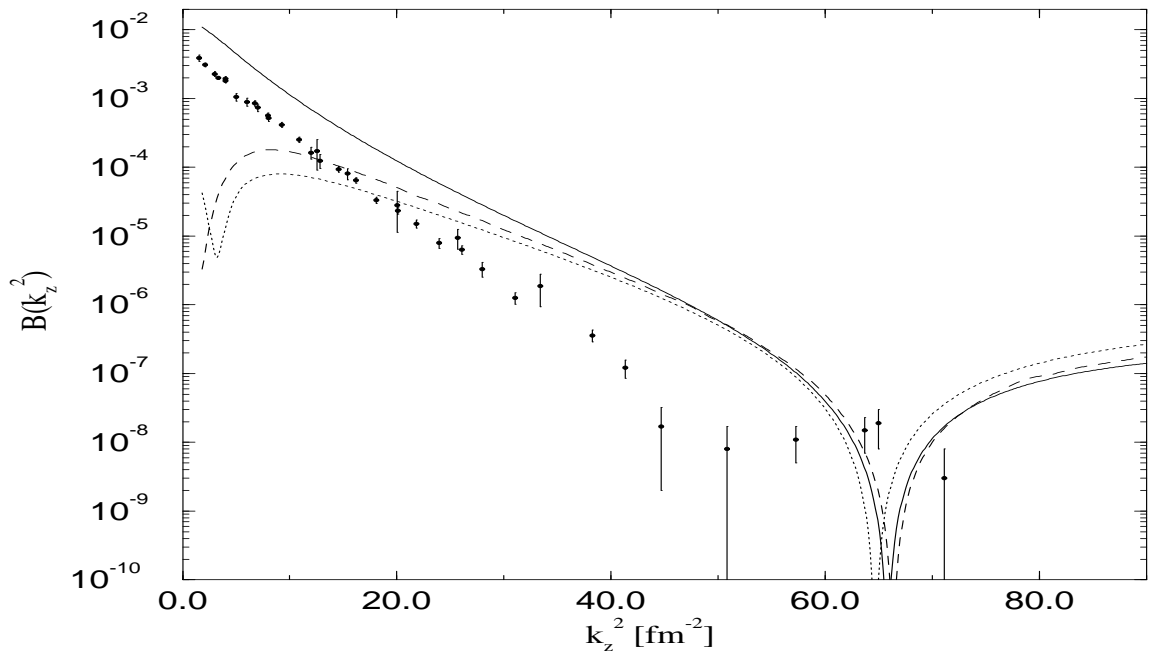}}
\caption{As Fig.\ 3, however for the formfactor $B(k_z^2)$ from equ.\ (64) \label{fffb}} 
\end{figure}

\begin{figure}[p]
\epsfxsize=  17.0cm
\epsfysize=  10.0cm
\centerline{\epsffile{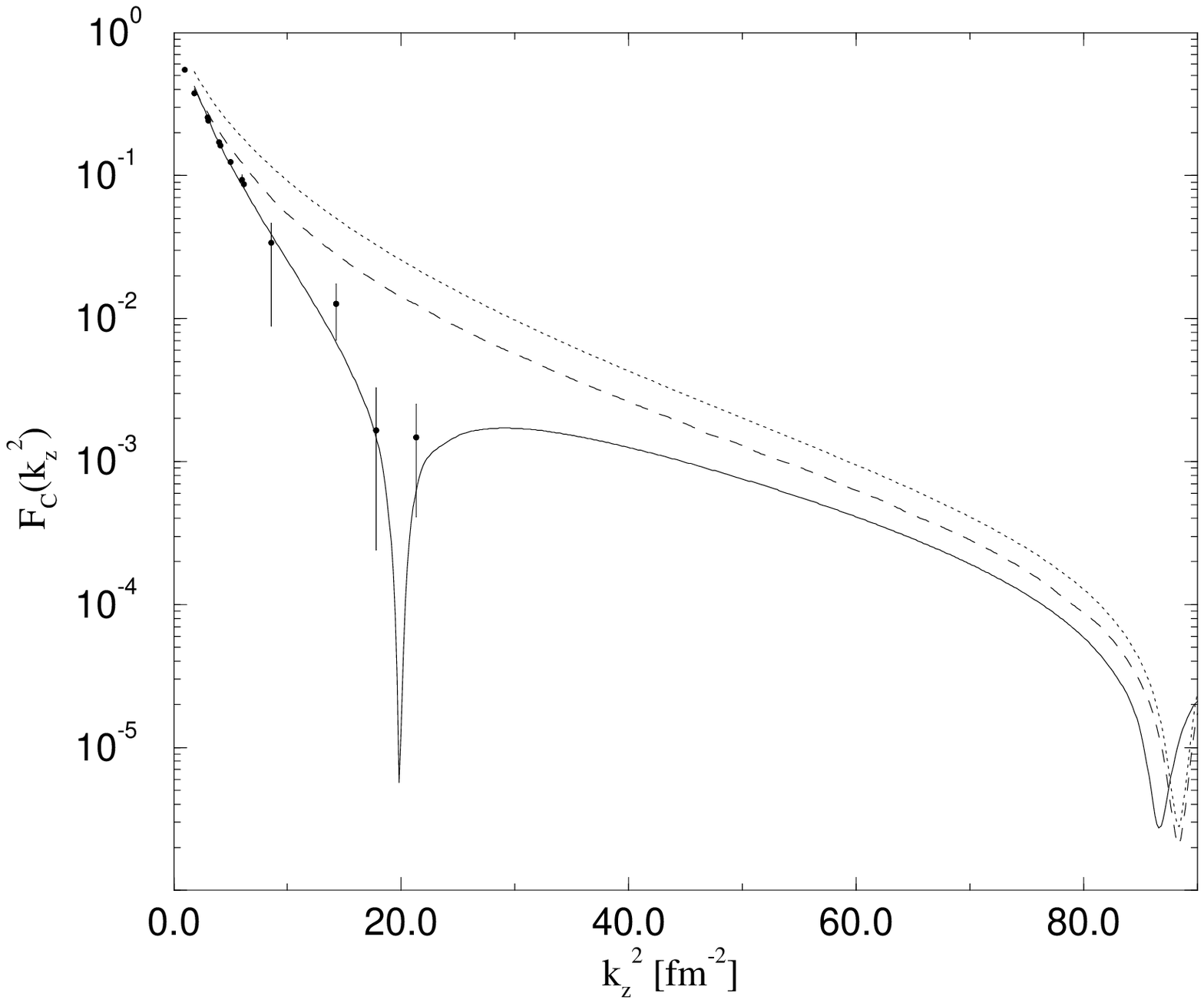}}
\caption{As Fig.\ 3, however for the formfactor $F_C(k_z^2)$ from equ.\ (57) \label{fffc}} 
\end{figure}

\begin{figure}[p]
\epsfxsize=  17.0cm
\epsfysize=  10.0cm
\centerline{\epsffile{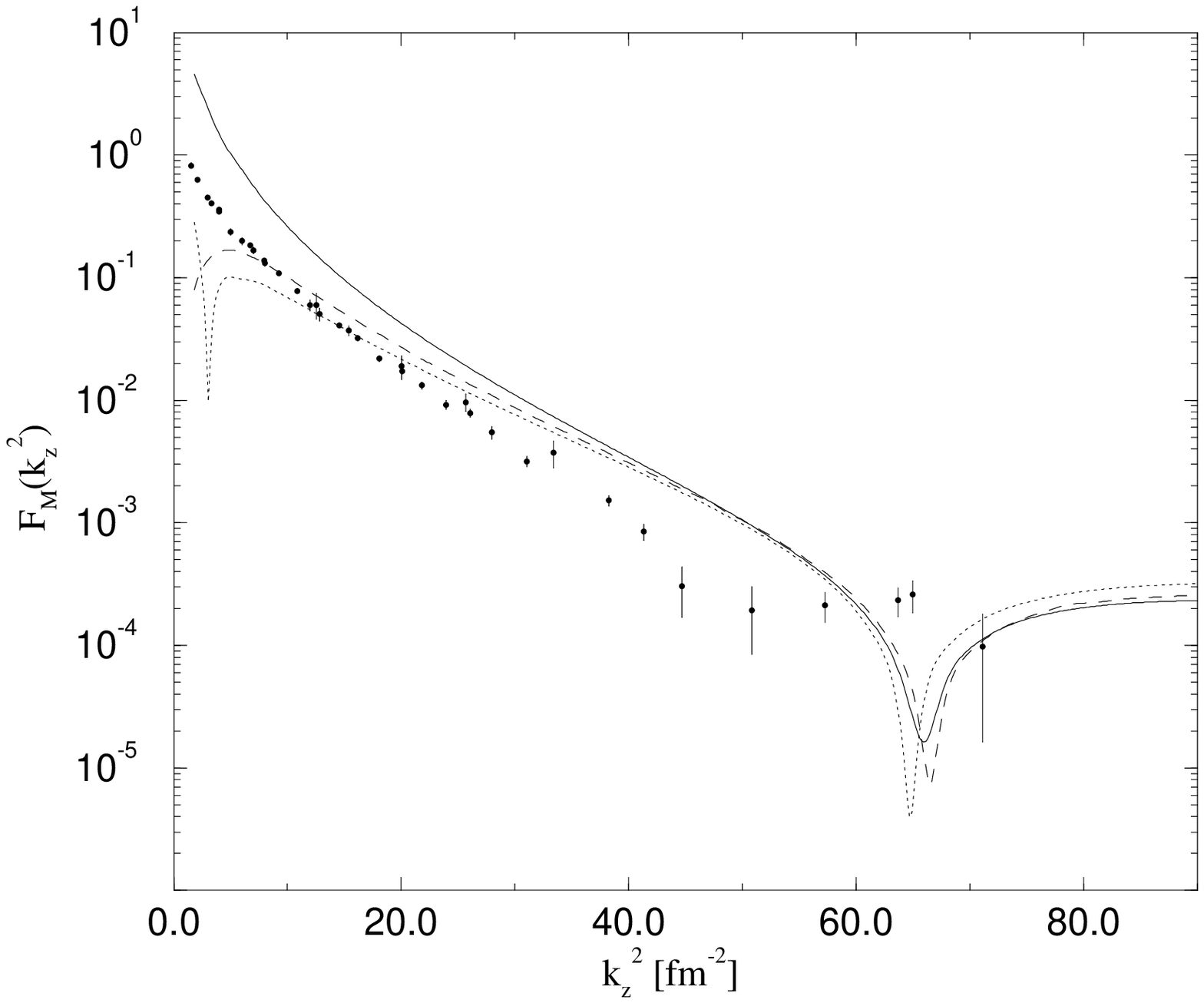}}
\caption{As Fig.\ 3, however for the formfactor $F_M(k_z^2)$ from equ.\ (58) \label{fffm}} 
\end{figure}

\begin{figure}[p]
\epsfxsize=  17.0cm
\epsfysize=  10.0cm
\centerline{\epsffile{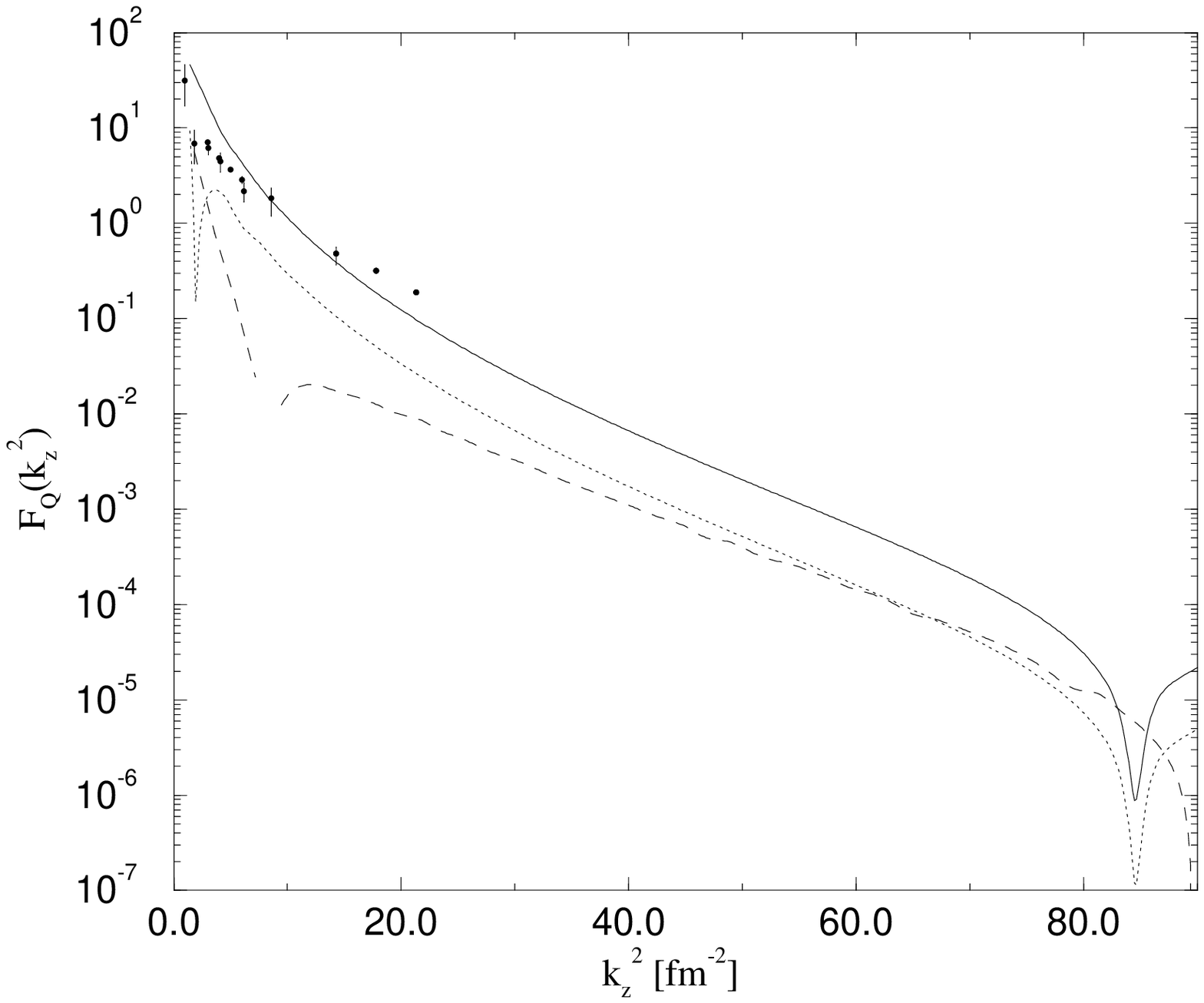}}
\caption{As Fig.\ 3, however for the formfactor $F_Q(k_z^2)$ from equ.\ (59) \label{fffq}} 
\end{figure}

\begin{figure}[p]
\epsfxsize=  17.0cm
\epsfysize=  10.0cm
\centerline{\epsffile{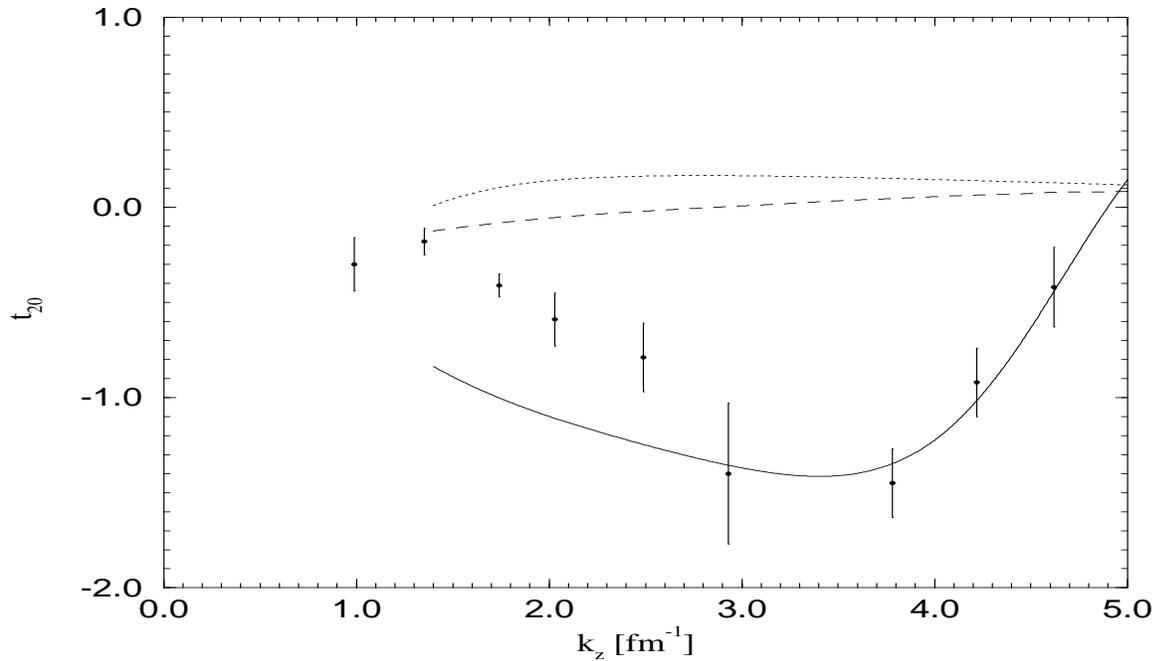}}
\caption{As Fig.\ 3, however for the tensor polarization $\tilde{t}_{\,20}(k_z)$ from equ.\ (65) \label{ffft}} 
\end{figure}

\begin{figure}[p]
\epsfxsize=  17.0cm
\epsfysize=  10.0cm
\centerline{\epsffile{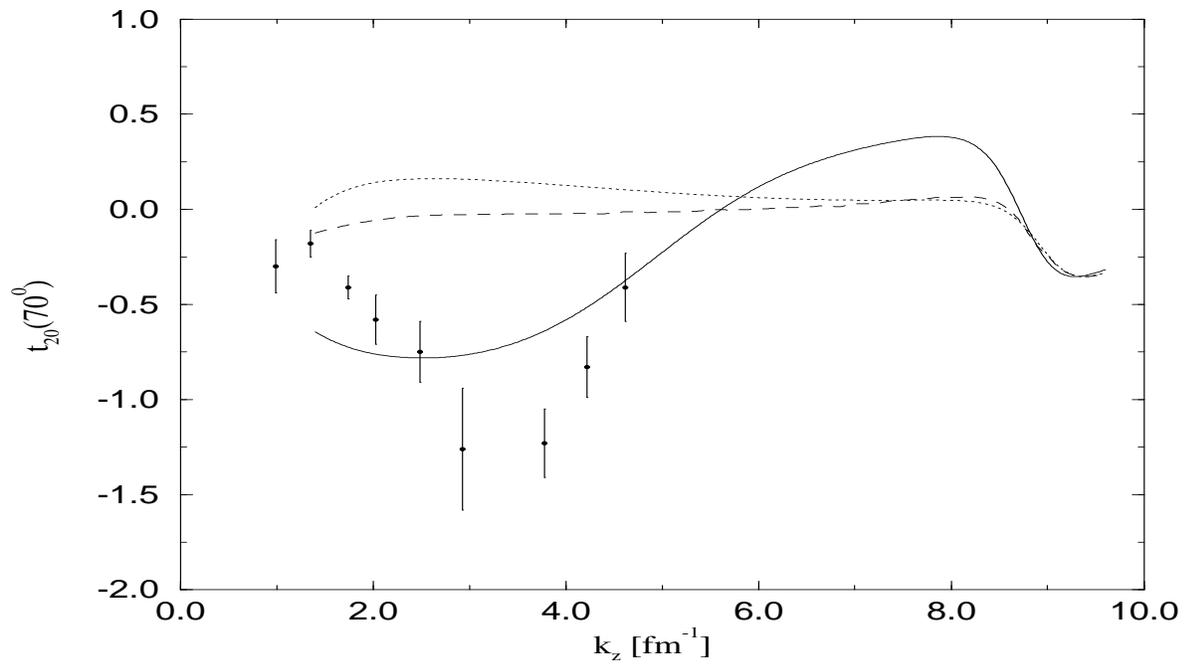}}
\caption{As Fig.\ 3, however for the tensor polarization $t_{\,20} \,(k_z,\theta_e=70^{\circ})$ from equ.\ (66) \label{fff7}} 
\end{figure}

As a characteristic result, formfactors and tensor polarizations of one selected
eigensolution of the BSE are shown in Figs.\ \ref{fffa} to \ref{fff7} for the  
set of parameters summarized in Table \ref{ttbb1}.

\begin{table}
\beqa \makebox[2.5cm][l]{-----} & & {\Lambda}_s\:=\:{\Lambda}_v\:={\Lambda}_p\:=\:0.24 , \nonumber \\
 & & E_B=2.371\cdot 10^{-3} \quad , \quad g_v=1.285714 \quad , \quad g_0=1. \quad , \quad g_p=0.897 \nonumber \\
 & & \Rightarrow \quad {\mu}_d = 6.685 \,{\mu}_K \quad , \quad Q_d = 1.287 \,{\mbox{fm}}^2 \nonumber \\
      \makebox[2.5cm][l]{$\cdot \cdot \cdot \cdot $} & & {\Lambda}_s\:=\:{\Lambda}_v\:={\Lambda}_p\:=\:0.24 , \nonumber \\
 & & E_B=2.371\cdot 10^{-3} \quad , \quad g_v=1.097139 \quad , \quad g_0=1. \quad , \quad g_p=1.522057 \nonumber \\
 & & \Rightarrow \quad {\mu}_d = 0.8570 \,{\mu}_K \quad , \quad Q_d = 0.2860 \,{\mbox{fm}}^2 \nonumber \\
      \makebox[2.5cm][l]{- - -} & & {\Lambda}_s\:=\:0.30 \quad , \quad {\Lambda}_v\:={\Lambda}_p\:=\:0.24 , \nonumber \\
 & & E_B=2.371\cdot 10^{-3} \quad , \quad g_v=2.309181 \quad , \quad g_0=1. \quad , \quad g_p=2.608756 \nonumber \\
 & & \Rightarrow \quad {\mu}_d = 0.8571 \,{\mu}_K \quad , \quad Q_d = 0.2860 \,{\mbox{fm}}^2 \nonumber 
\eeqa 

\caption{Parameters used in Figs.\ 3 to 9 (Interaction parameters and binding
energy are given in units of the nucleon mass, coupling constants are dimensionless, 
${\mu}_K = \frac{e}{2m}$ is the nuclear magneton) \label{ttbb1}}
\end{table}

To get a slight feeling for the dependence of observables like ${\mu}_d$ and
$Q_d$ on the parameters of our model we fixed the first
node of the charge formfactor $F_C$ to about $k_z^2=20 \,{\mbox{fm}}^{-2}$ (which is
suggested by experiment) and varied coupling constants and interaction parameters.
Typical results are listed in Table \ref{ttbb2}.

\begin{table}

\begin{tabular}{|l|l|l|l|l||l|l|}
   {${\Lambda}_s$}
 & {${\Lambda}_v$}
 & {${\Lambda}_p$}
 & {$g_v$}
 & {$g_p$}
 & {${\mu}_d/{\mu}_K$}
 & {$Q_d$ [${\mbox{fm}}^{2}$] } \\ \hline \hline
 0.24 & 0.24 & 0.22 & 1.285714 & 0.89     & 6.413 & 1.261 \\
 0.24 & 0.24 & 0.24 & 1.285714 & 0.897    & 6.685 & 1.287 \\
 0.24 & 0.24 & 0.24 & 1.285714 & 2.85     & 6.674 & 1.340 \\
 0.32 & 0.32 & 0.32 & 1.171429 & 2.5      & 7.230 & 1.102 \\
 0.32 & 0.32 & 0.32 & 1.285714 & 2.1      & 6.977 & 0.7958 \\
 0.32 & 0.32 & 0.32 & 3.0      & 2.55     & 6.980 & 0.6127 \\
 0.32 & 0.32 & 0.32 & 9.142858 & 3.540875 & 6.973 & 0.7586 \\
\end{tabular}

\caption{Typical parameter dependence of the magnetic
and quadrupole moment for $g_0=1$ and $E_B=2.371\cdot 10^{-3}$ \label{ttbb2}}
\end{table}

Clearly, at present our parameter studies are by far not
exhaustive due to practical restrictions. For unequal interaction parameters the analytical and 
numerical integrations for the boost of BSAs from rest
frame of the deuteron to the Breit frame are very involved and 
time expensive, additionally the positive parity eigensolutions 
of the BSE in our model are no more degenerate. Finally fixing the node
of the charge formfactor to $k_z^2=20 \,{\mbox{fm}}^{-2}$ is only achieved
by a very fine tuning of the coupling constants. All this 
makes a systematic discussion
of the coupling space of our model extremly difficult. Nevertheless, by
our experience with the model we draw some more general statements
to our results in the conclusions.

We close with a final remark. It has been shown by B.\ Michel \cite{mic1} that for ${\Lambda}_s={\Lambda}_v={\Lambda}_p$
and $g_0=1$ positive parity eigensolutions of the model can be controlled 
by one simple parameter $y_{-}$, which is defined by:

\beq y_{-} \, = \, 
\frac{[g_p + (4 J - 3) \, g_v] + (1 - g_v) \, \frac{\displaystyle
X^{--(J)}}{\displaystyle X^{++(J)}}}{(1 - g_v) + [g_p + (4 J - 3) \, g_v] \, \frac{\displaystyle X^{--(J)}}{\displaystyle X^{++(J)}}} \eeq

and which is the ratio between the $++$ and the $--$ component of the BSA (because of the 
degeneracy of the interaction parameters the index $j$ of $X^{++(J)}_j$ and $X^{--(J)}_j$ is
dropped). We defer a more detailed discussion of this interesting feature to a
forthcoming publication.

\section{Conclusion}

In this paper we developped the formalism for the  covariant description of
bound fermion--antifermion systems in the framework of the BSE. To faciliate
the very complex solution of the problem, the kernel of BSE was represented by
a covariant one--rank separable interaction piece for each of our 3
spin--invariants, which reduces the solution of the BSE to an algebraic
problem. By integration over the relative energy variable, the full nonstatic
BSAs were related to standard static 3--dimensional wave functions in momentum
space. \\

As a first step we applied our formalism to the deuteron and
investigated frame independent, i.e.\ covariant "deuteron" wavefuctions
together with the corresponding IA formfactors and tensor polarizations.
The normalization of the BSA was obtained by normalizing the charge formfactor
to 1 for zero momentum transfer. We find that our simple interaction kernel is obviously not able to describe the deuteron
accurately. Explicitly there are two scenarios: if, on the one hand, we reproduce magnetic and quadrupole
moments, we fail to reproduce the $k^2$ dependence of the formfactors. On the other
hand, upon fixing the $k^2$ dependence qualitatively 
from the first node of the charge formfactor $F_C$ at about $k_z^2=20 \,{\mbox{fm}}^{-2}$, the moments turn out to be
too large. Nevertheless, it is interesting to see that we are able to control
the ${}^3D^+_1$ wave and the antiparticle content of our BSA over a wide range
(i.e.\ it is no problem to obtain ${}^3D^+_1$ wave admixtures of 5 \%, even without
an explicit tensor force in our
interaction kernel).

From its ansatz, our approach is just a first, crude step towards a more
realistic covariant description
of relativistic bound systems. The crucial point is certainly a more adequate formulation of the
interaction kernel within a systematic separable expansion. Such an extension,
which is presently under way, then opens up a variety of
interesting questions within the model, to name only a systematic investigation of mesonic systems in
standard coordinates and on the light cone.
\newpage
\begin{figure}[p]
\unitlength1cm
\begin{picture}(17,10)
\thinlines
\end{picture}
\end{figure}
\clearpage
%
% Anhang:
% =======                              
%

\begin{appendix}
\section{The partial waves of the ETW in the rest frame of the deuteron} \label{app1}
Using the definitions

\beq {\tilde{\omega}}^{\,\pm \,}(\vec{q}) \quad := \quad \omega (\vec{q}) \pm m \qquad , \qquad {\tilde{\omega}}^{\epm}(\vec{q}) \quad := \quad \omega (\vec{q}) \xpm m \eeq

we give here the expicit expressions for the partial waves
obtained by evaluation of equation (\ref{ppp1}).

We get for $J=0$:

\beqa
 {\Phi}_{\displaystyle {\,}^1S_0^{+}} \, (M_d\, ; \oabs{q}) & = & 
 -\,\sum\limits_{\pm \epm } \frac{\sqrt{\pi}}{2}\:\cdot \:\frac{(g^{(0)}(M_d\, ,\vec{0}))^2}{(2\pi )^3\,{\omega}^{\,2} (\vec{q})} \quad \cdot \nonumber \\
 & & \nonumber \\
 & \cdot \; \Biggl\{ \Biggr. & w_{_{(s)}}^{\pm \epm } (E_N ; \vec{q}) \: ({\tilde{\omega}}^{\,\pm \,} (\vec{q}) \,{\tilde{\omega}}^{\,\epm \,} (\vec{q}) \:X_s^{++(0)} \pm \xpm \:X_s^{--(0)} \cdot \oabsq{q}) \quad + \nonumber \\
 & & \nonumber \\
 & + & w_{_{(v)}}^{\pm \epm } (E_N ; \vec{q}) \:g_v \: \Bigl( - \,({\tilde{\omega}}^{\,\pm \,} (\vec{q}) \, {\tilde{\omega}}^{\,\epm \,} (\vec{q}) \pm \xpm 3\,g_0\,\oabsq{q})\:X_v^{++(0)} \quad - \nonumber \\
 & & \nonumber \\
 & - & (3\,g_0\, {\tilde{\omega}}^{\,\pm \,} (\vec{q}) \, {\tilde{\omega}}^{\,\epm \,} (\vec{q}) \pm \xpm \oabsq{q}) \:X_v^{--(0)} \Bigr) \quad + \nonumber \\
 & & \nonumber \\
 & + & w_{_{(p)}}^{\pm \epm } (E_N ; \vec{q}) \:g_p \: ({\tilde{\omega}}^{\,\pm \,} (\vec{q}) \, {\tilde{\omega}}^{\,\epm \,} (\vec{q}) \:X_p^{--(0)} \pm \xpm \:X_p^{++(0)} \cdot \oabsq{q}) \; \Biggr\} \qquad \nonumber
\eeqa

\beqa
 {\Phi}_{\displaystyle {\,}^3P_0^{+}} \, (M_d\, ; \oabs{q}) & = & 
 \sum\limits_{\pm \epm } \frac{\sqrt{\pi}}{2}\:\cdot \:\frac{(g^{(0)}(M_d\, ,\vec{0}))^2}{(2\pi )^3\,{\omega}^{\,2} (\vec{q})} \: \cdot \: \oabs{q} \: \cdot \nonumber \\
 & & \nonumber \\
 & \cdot \; \Biggl\{ \Biggr. & w_{_{(s)}}^{\pm \epm } (E_N ; \vec{q}) \: (\pm \, {\tilde{\omega}}^{\,\emp \,} (\vec{q}) \:X_s^{--(0)} \xpm {\tilde{\omega}}^{\,\pm \,} (\vec{q}) \,\:X_s^{++(0)} ) \quad + \nonumber \\
 & & \nonumber \\
 & + & w_{_{(v)}}^{\pm \epm } (E_N ; \vec{q}) \:g_v \: \Bigl( - \,(\pm \, 3\,g_0 \, {\tilde{\omega}}^{\,\emp \,} (\vec{q}) \xpm {\tilde{\omega}}^{\,\pm \,} (\vec{q}) )\:X_v^{++(0)} \quad - \nonumber \\
 & & \nonumber \\
 & - & (\pm \, {\tilde{\omega}}^{\,\emp \,} (\vec{q}) \xpm 3\,g_0\,{\tilde{\omega}}^{\,\pm \,} (\vec{q}) ) \:X_v^{--(0)} \Bigr) \quad + \nonumber \\
 & & \nonumber \\
 & + & w_{_{(p)}}^{\pm \epm } (E_N ; \vec{q}) \:g_p \: (\pm \, {\tilde{\omega}}^{\,\emp \,} (\vec{q}) \:X_p^{++(0)} \xpm {\tilde{\omega}}^{\,\pm \,} (\vec{q}) \, \:X_p^{--(0)} ) \Biggr\} \qquad \nonumber
\eeqa

\beqa
 {\Phi}_{\displaystyle {\,}^3P_0^{-}} \, (M_d\, ; \oabs{q}) & = & 
 \sum\limits_{\pm \epm } \frac{\sqrt{\pi}}{2}\:\cdot \:\frac{(g^{(0)}(M_d\, ,\vec{0}))^2}{(2\pi )^3\,{\omega}^{\,2} (\vec{q})} \: \cdot \: \oabs{q} \: \cdot \nonumber \\
 & & \nonumber \\
 & \cdot \; \Biggl\{ \Biggr. & w_{_{(s)}}^{\pm \epm } (E_N ; \vec{q}) \: (\pm \, {\tilde{\omega}}^{\,\epm \,} (\vec{q}) \:X_s^{++(0)} \xpm {\tilde{\omega}}^{\,\mp \,} (\vec{q}) \,\:X_s^{--(0)} ) \quad + \nonumber \\
 & & \nonumber \\
 & + & w_{_{(v)}}^{\pm \epm } (E_N ; \vec{q}) \:g_v \: \Bigl( - \,(\pm \, 3\,g_0 \, {\tilde{\omega}}^{\,\epm \,} (\vec{q}) \xpm {\tilde{\omega}}^{\,\mp \,} (\vec{q}) )\:X_v^{--(0)} \quad - \nonumber \\
 & & \nonumber \\
 & - & (\pm \, {\tilde{\omega}}^{\,\epm \,} (\vec{q}) \xpm 3\,g_0\,{\tilde{\omega}}^{\,\mp \,} (\vec{q}) ) \:X_v^{++(0)} \Bigr) \quad + \nonumber \\
 & & \nonumber \\
 & + & w_{_{(p)}}^{\pm \epm } (E_N ; \vec{q}) \:g_p \: (\pm \, {\tilde{\omega}}^{\,\epm \,} (\vec{q}) \:X_p^{--(0)} \xpm {\tilde{\omega}}^{\,\mp \,} (\vec{q}) \, \:X_p^{++(0)} ) \Biggr\} \qquad \nonumber
\eeqa

\beqa
 {\Phi}_{\displaystyle {\,}^1S_0^{-}} \, (M_d\, ; \oabs{q}) & = & 
 -\,\sum\limits_{\pm \epm } \frac{\sqrt{\pi}}{2}\:\cdot \:\frac{(g^{(0)}(M_d\, ,\vec{0}))^2}{(2\pi )^3\,{\omega}^{\,2} (\vec{q})} \quad \cdot \nonumber \\
 & & \nonumber \\
 & \cdot \; \Biggl\{ \Biggr. & w_{_{(s)}}^{\pm \epm } (E_N ; \vec{q}) \: ({\tilde{\omega}}^{\,\mp \,} (\vec{q}) \,{\tilde{\omega}}^{\,\emp \,} (\vec{q}) \:X_s^{--(0)} \pm \xpm \:X_s^{++(0)} \cdot \oabsq{q}) \quad + \nonumber \\
 & & \nonumber \\
 & + & w_{_{(v)}}^{\pm \epm } (E_N ; \vec{q}) \:g_v \: \Bigl( - \,({\tilde{\omega}}^{\,\mp \,} (\vec{q}) \, {\tilde{\omega}}^{\,\emp \,} (\vec{q}) \pm \xpm 3\,g_0\,\oabsq{q})\:X_v^{--(0)} \quad - \nonumber \\
 & & \nonumber \\
 & - & (3\,g_0\, {\tilde{\omega}}^{\,\mp \,} (\vec{q}) \, {\tilde{\omega}}^{\,\emp \,} (\vec{q}) \pm \xpm \oabsq{q}) \:X_v^{++(0)} \Bigr) \quad + \nonumber \\
 & & \nonumber \\
 & + & w_{_{(p)}}^{\pm \epm } (E_N ; \vec{q}) \:g_p \: ({\tilde{\omega}}^{\,\mp \,} (\vec{q}) \, {\tilde{\omega}}^{\,\emp \,} (\vec{q}) \:X_p^{++(0)} \pm \xpm \:X_p^{--(0)} \cdot \oabsq{q}) \; \Biggr\} \qquad \nonumber
\eeqa

and for $J=1$:

\beqa
 {\Phi}_{\displaystyle {\,}^3S_1^{+}} \, (M_d\, ; \oabs{q}) & = & 
 -\,\sum\limits_{\pm \epm } \frac{\sqrt{\pi}}{2}\:\cdot \:\frac{(g^{(1)}(M_d\, ,\vec{0}))^2}{(2\pi )^3\,{\omega}^{\,2} (\vec{q})} \quad \cdot \nonumber \\
 & & \nonumber \\
 & \cdot \; \Biggl\{ \Biggr. & w_{_{(s)}}^{\pm \epm } (E_N ; \vec{q}) \: ({\tilde{\omega}}^{\,\pm \,} (\vec{q}) \,{\tilde{\omega}}^{\,\epm \,} (\vec{q}) \:X_s^{++(1)} \pm \xpm (- \,\frac{1}{3} ) \:X_s^{--(1)} \cdot \oabsq{q}) \quad + \nonumber \\
 & & \nonumber \\
 & + & w_{_{(v)}}^{\pm \epm } (E_N ; \vec{q}) \:g_v \: \Bigl( (- \,{\tilde{\omega}}^{\,\pm \,} (\vec{q}) \, {\tilde{\omega}}^{\,\epm \,} (\vec{q}) \pm \xpm (- \,\frac{1}{3} )\,g_0\,\oabsq{q})\:X_v^{++(1)} \quad + \nonumber \\
 & & \nonumber \\
 & + & (g_0\, {\tilde{\omega}}^{\,\pm \,} (\vec{q}) \, {\tilde{\omega}}^{\,\epm \,} (\vec{q}) \pm \xpm \,\frac{1}{3} \, \oabsq{q}) \:X_v^{--(1)} \Bigr) \quad + \nonumber \\
 & & \nonumber \\
 & + & w_{_{(p)}}^{\pm \epm } (E_N ; \vec{q}) \,g_p \, ({\tilde{\omega}}^{\,\pm \,} (\vec{q}) \, {\tilde{\omega}}^{\,\epm \,} (\vec{q}) \,X_p^{--(1)} \pm \xpm (- \,\frac{1}{3} )\,X_p^{++(1)} \cdot \oabsq{q}) \! \Biggr\} \, \nonumber
\eeqa

\beqa
 {\Phi}_{\displaystyle {\,}^3D_1^{+}} \, (M_d\, ; \oabs{q}) & = & 
 \sum\limits_{\pm \epm } \pm \xpm \frac{\sqrt{2\,\pi}}{3}\:\cdot \:\frac{(g^{(1)}(M_d\, ,\vec{0}))^2}{(2\pi )^3\,{\omega}^{\,2} (\vec{q})} \: \cdot \: \oabsq{q} \: \cdot 
 \; \Biggl\{ \Biggr. w_{_{(s)}}^{\pm \epm } (E_N ; \vec{q}) \: X_s^{--(1)} \quad + \nonumber \\
 & & \nonumber \\
 & + & w_{_{(v)}}^{\pm \epm } (E_N ; \vec{q}) \:g_v \: (-\, X_v^{--(1)} + \,g_0\,X_v^{++(1)} ) \quad + \nonumber \\
 & & \nonumber \\
 & + & w_{_{(p)}}^{\pm \epm } (E_N ; \vec{q}) \:g_p \: X_p^{++(1)} \Biggr\} \qquad \nonumber
\eeqa

\beqa
 {\Phi}_{\displaystyle {\,}^3P_1^{+}} \, (M_d\, ; \oabs{q}) & = & 
 \sum\limits_{\pm \epm } \sqrt{\frac{\pi}{6}}\:\cdot \:\frac{(g^{(1)}(M_d\, ,\vec{0}))^2}{(2\pi )^3\,{\omega}^{\,2} (\vec{q})} \: \cdot \: \oabs{q} \: \cdot \nonumber \\
 & & \nonumber \\
 & \cdot \; \Biggl\{ \Biggr. & w_{_{(s)}}^{\pm \epm } (E_N ; \vec{q}) \: (\pm \, {\tilde{\omega}}^{\,\emp \,} (\vec{q}) \:X_s^{--(1)} - \xpm {\tilde{\omega}}^{\,\pm \,} (\vec{q}) \,\:X_s^{++(1)} ) \quad + \nonumber \\
 & & \nonumber \\
 & + & w_{_{(v)}}^{\pm \epm } (E_N ; \vec{q}) \:g_v \: \Bigl( (\pm \, g_0 \, {\tilde{\omega}}^{\,\emp \,} (\vec{q}) \xpm {\tilde{\omega}}^{\,\pm \,} (\vec{q}) )\:X_v^{++(1)} \quad - \nonumber \\
 & & \nonumber \\
 & - & (\pm \, {\tilde{\omega}}^{\,\emp \,} (\vec{q}) \xpm \,g_0\,{\tilde{\omega}}^{\,\pm \,} (\vec{q}) ) \:X_v^{--(1)} \Bigr) \quad + \nonumber \\
 & & \nonumber \\
 & + & w_{_{(p)}}^{\pm \epm } (E_N ; \vec{q}) \:g_p \: (\pm \, {\tilde{\omega}}^{\,\emp \,} (\vec{q}) \:X_p^{++(1)} - \xpm {\tilde{\omega}}^{\,\pm \,} (\vec{q}) \, \:X_p^{--(1)} ) \Biggr\} \qquad \nonumber
\eeqa

\beqa
 {\Phi}_{\displaystyle {\,}^1P_1^{+}} \, (M_d\, ; \oabs{q}) & = & 
 -\, \sum\limits_{\pm \epm } \frac{\sqrt{\pi}}{2\,\sqrt{3}}\:\cdot \:\frac{(g^{(1)}(M_d\, ,\vec{0}))^2}{(2\pi )^3\,{\omega}^{\,2} (\vec{q})} \: \cdot \: \oabs{q} \: \cdot \nonumber \\
 & & \nonumber \\
 & \cdot \; \Biggl\{ \Biggr. & w_{_{(s)}}^{\pm \epm } (E_N ; \vec{q}) \: (\pm \, {\tilde{\omega}}^{\,\emp \,} (\vec{q}) \:X_s^{--(1)} \xpm {\tilde{\omega}}^{\,\pm \,} (\vec{q}) \,\:X_s^{++(1)} ) \quad + \nonumber \\
 & & \nonumber \\
 & + & w_{_{(v)}}^{\pm \epm } (E_N ; \vec{q}) \:g_v \: \Bigl( (\pm \, g_0 \, {\tilde{\omega}}^{\,\emp \,} (\vec{q}) - \xpm {\tilde{\omega}}^{\,\pm \,} (\vec{q}) )\:X_v^{++(1)} \quad - \nonumber \\
 & & \nonumber \\
 & - & (\pm \, {\tilde{\omega}}^{\,\emp \,} (\vec{q}) - \xpm \,g_0\,{\tilde{\omega}}^{\,\pm \,} (\vec{q}) ) \:X_v^{--(1)} \Bigr) \quad + \nonumber \\
 & & \nonumber \\
 & + & w_{_{(p)}}^{\pm \epm } (E_N ; \vec{q}) \:g_p \: (\pm \, {\tilde{\omega}}^{\,\emp \,} (\vec{q}) \:X_p^{++(1)} \xpm {\tilde{\omega}}^{\,\pm \,} (\vec{q}) \, \:X_p^{--(1)} ) \Biggr\} \qquad \nonumber
\eeqa

\beqa
 {\Phi}_{\displaystyle {\,}^3P_1^{-}} \, (M_d\, ; \oabs{q}) & = & 
 \sum\limits_{\pm \epm } \sqrt{\frac{\pi}{6}}\:\cdot \:\frac{(g^{(1)}(M_d\, ,\vec{0}))^2}{(2\pi )^3\,{\omega}^{\,2} (\vec{q})} \: \cdot \: \oabs{q} \: \cdot \nonumber \\
 & & \nonumber \\
 & \cdot \; \Biggl\{ \Biggr. & w_{_{(s)}}^{\pm \epm } (E_N ; \vec{q}) \: (\pm \, {\tilde{\omega}}^{\,\epm \,} (\vec{q}) \:X_s^{++(1)} - \xpm {\tilde{\omega}}^{\,\mp \,} (\vec{q}) \,\:X_s^{--(1)} ) \quad + \nonumber \\
 & & \nonumber \\
 & + & w_{_{(v)}}^{\pm \epm } (E_N ; \vec{q}) \:g_v \: \Bigl( (\pm \, g_0 \, {\tilde{\omega}}^{\,\epm \,} (\vec{q}) \xpm {\tilde{\omega}}^{\,\mp \,} (\vec{q}) )\:X_v^{--(1)} \quad - \nonumber \\
 & & \nonumber \\
 & - & (\pm \, {\tilde{\omega}}^{\,\epm \,} (\vec{q}) \xpm \,g_0\,{\tilde{\omega}}^{\,\mp \,} (\vec{q}) ) \:X_v^{++(1)} \Bigr) \quad + \nonumber \\
 & & \nonumber \\
 & + & w_{_{(p)}}^{\pm \epm } (E_N ; \vec{q}) \:g_p \: (\pm \, {\tilde{\omega}}^{\,\epm \,} (\vec{q}) \:X_p^{--(1)} - \xpm {\tilde{\omega}}^{\,\mp \,} (\vec{q}) \, \:X_p^{++(1)} ) \Biggr\} \qquad \nonumber
\eeqa

\beqa
 {\Phi}_{\displaystyle {\,}^1P_1^{-}} \, (M_d\, ; \oabs{q}) & = & 
 - \,\sum\limits_{\pm \epm } \frac{\sqrt{\pi}}{2\,\sqrt{3}}\:\cdot \:\frac{(g^{(1)}(M_d\, ,\vec{0}))^2}{(2\pi )^3\,{\omega}^{\,2} (\vec{q})} \: \cdot \: \oabs{q} \: \cdot \nonumber \\
 & & \nonumber \\
 & \cdot \; \Biggl\{ \Biggr. & w_{_{(s)}}^{\pm \epm } (E_N ; \vec{q}) \: (\pm \, {\tilde{\omega}}^{\,\epm \,} (\vec{q}) \:X_s^{++(1)} \xpm {\tilde{\omega}}^{\,\mp \,} (\vec{q}) \,\:X_s^{--(1)} ) \quad + \nonumber \\
 & & \nonumber \\
 & + & w_{_{(v)}}^{\pm \epm } (E_N ; \vec{q}) \:g_v \: \Bigl( (\pm \, g_0 \, {\tilde{\omega}}^{\,\epm \,} (\vec{q}) - \xpm {\tilde{\omega}}^{\,\mp \,} (\vec{q}) )\:X_v^{--(1)} \quad - \nonumber \\
 & & \nonumber \\
 & - & (\pm \, {\tilde{\omega}}^{\,\epm \,} (\vec{q}) - \xpm \,g_0\,{\tilde{\omega}}^{\,\mp \,} (\vec{q}) ) \:X_v^{++(1)} \Bigr) \quad + \nonumber \\
 & & \nonumber \\
 & + & w_{_{(p)}}^{\pm \epm } (E_N ; \vec{q}) \:g_p \: (\pm \, {\tilde{\omega}}^{\,\epm \,} (\vec{q}) \:X_p^{--(1)} \xpm {\tilde{\omega}}^{\,\mp \,} (\vec{q}) \, \:X_p^{++(1)} ) \Biggr\} \qquad \nonumber
\eeqa

\beqa
 {\Phi}_{\displaystyle {\,}^3S_1^{-}} \, (M_d\, ; \oabs{q}) & = & 
 -\,\sum\limits_{\pm \epm } \frac{\sqrt{\pi}}{2}\:\cdot \:\frac{(g^{(1)}(M_d\, ,\vec{0}))^2}{(2\pi )^3\,{\omega}^{\,2} (\vec{q})} \quad \cdot \nonumber \\
 & & \nonumber \\
 & \cdot \; \Biggl\{ \Biggr. & w_{_{(s)}}^{\pm \epm } (E_N ; \vec{q}) \: ({\tilde{\omega}}^{\,\mp \,} (\vec{q}) \,{\tilde{\omega}}^{\,\emp \,} (\vec{q}) \:X_s^{--(1)} \pm \xpm (- \,\frac{1}{3} ) \:X_s^{++(1)} \cdot \oabsq{q}) \quad + \nonumber \\
 & & \nonumber \\
 & + & w_{_{(v)}}^{\pm \epm } (E_N ; \vec{q}) \:g_v \: \Bigl( (- \,{\tilde{\omega}}^{\,\mp \,} (\vec{q}) \, {\tilde{\omega}}^{\,\emp \,} (\vec{q}) \pm \xpm (- \,\frac{1}{3} )\,g_0\,\oabsq{q})\:X_v^{--(1)} \quad + \nonumber \\
 & & \nonumber \\
 & + & (g_0\, {\tilde{\omega}}^{\,\mp \,} (\vec{q}) \, {\tilde{\omega}}^{\,\emp \,} (\vec{q}) \pm \xpm \,\frac{1}{3} \, \oabsq{q}) \:X_v^{++(1)} \Bigr) \quad + \nonumber \\
 & & \nonumber \\
 & + & w_{_{(p)}}^{\pm \epm } (E_N ; \vec{q}) \,g_p \, ({\tilde{\omega}}^{\,\mp \,} (\vec{q}) \, {\tilde{\omega}}^{\,\emp \,} (\vec{q}) \,X_p^{++(1)} \pm \xpm (- \,\frac{1}{3} )\,X_p^{--(1)} \cdot \oabsq{q}) \! \Biggr\} \, \nonumber
\eeqa

\beqa
 {\Phi}_{\displaystyle {\,}^3D_1^{-}} \, (M_d\, ; \oabs{q}) & = & 
 \sum\limits_{\pm \epm } \pm \xpm \frac{\sqrt{2\,\pi}}{3}\:\cdot \:\frac{(g^{(1)}(M_d\, ,\vec{0}))^2}{(2\pi )^3\,{\omega}^{\,2} (\vec{q})} \: \cdot \: \oabsq{q} \: \cdot 
 \; \Biggl\{ \Biggr. w_{_{(s)}}^{\pm \epm } (E_N ; \vec{q}) \: X_s^{++(1)} \quad + \nonumber \\
 & & \nonumber \\
 & + & w_{_{(v)}}^{\pm \epm } (E_N ; \vec{q}) \:g_v \: (-\, X_v^{++(1)} + \,g_0\,X_v^{--(1)} ) \quad + \nonumber \\
 & & \nonumber \\
 & + & w_{_{(p)}}^{\pm \epm } (E_N ; \vec{q}) \:g_p \: X_p^{--(1)} \Biggr\} \qquad \nonumber
\eeqa
\section{List of experimental data} \label{app2}
The following tables consist of all experimental data we have used in our
plots. They are either original or derived data from the references quoted
at each line in the tables.

\begin{tabular}{rccl}
 & & & \\
 {$k_z^2$ [${\mbox{fm}}^{-2}$] }
 & {$B(k_z^2)$}
 & {$F_M(k_z^2)$}
 & Ref. \\ \hline \hline
 & & & \\
 {$\mbox{ .1550E+01}$}
 & {${\mbox{ .3900E-02}}^{\,+\mbox{ .4000E-03}}_{\,-\mbox{ .4000E-03}}$}
 & {${\mbox{ .8235E+00}}^{\,+\mbox{ .4223E-01}}_{\,-\mbox{ .4223E-01}}$}
 & Sim81 \\
 {$\mbox{ .2100E+01}$}
 & {${\mbox{ .3100E-02}}^{\,+\mbox{ .2000E-03}}_{\,-\mbox{ .2000E-03}}$}
 & {${\mbox{ .6303E+00}}^{\,+\mbox{ .2033E-01}}_{\,-\mbox{ .2033E-01}}$}
 & Sim81 \\
 {$\mbox{ .3000E+01}$}
 & {${\mbox{ .2273E-02}}^{\,+\mbox{ .1915E-03}}_{\,-\mbox{ .1915E-03}}$}
 & {${\mbox{ .4510E+00}}^{\,+\mbox{ .1900E-01}}_{\,-\mbox{ .1900E-01}}$}
 & Ben66 \\
 {$\mbox{ .3300E+01}$}
 & {${\mbox{ .2000E-02}}^{\,+\mbox{ .1000E-03}}_{\,-\mbox{ .1000E-03}}$}
 & {${\mbox{ .4032E+00}}^{\,+\mbox{ .1008E-01}}_{\,-\mbox{ .1008E-01}}$}
 & Sim81 \\
 {$\mbox{ .4000E+01}$}
 & {${\mbox{ .1936E-02}}^{\,+\mbox{ .1614E-03}}_{\,-\mbox{ .1614E-03}}$}
 & {${\mbox{ .3600E+00}}^{\,+\mbox{ .1500E-01}}_{\,-\mbox{ .1500E-01}}$}
 & Ben66 \\
 {$\mbox{ .4000E+01}$}
 & {${\mbox{ .1800E-02}}^{\,+\mbox{ .1000E-03}}_{\,-\mbox{ .1000E-03}}$}
 & {${\mbox{ .3471E+00}}^{\,+\mbox{ .9641E-02}}_{\,-\mbox{ .9641E-02}}$}
 & Sim81 \\
\end{tabular}

\begin{tabular}{rccl}
 {$k_z^2$ [${\mbox{fm}}^{-2}$] }
 & {$B(k_z^2)$}
 & {$F_M(k_z^2)$}
 & Ref. \\ \hline \hline
 & & & \\
 {$\mbox{ .5000E+01}$}
 & {${\mbox{ .1052E-02}}^{\,+\mbox{ .1332E-03}}_{\,-\mbox{ .1332E-03}}$}
 & {${\mbox{ .2370E+00}}^{\,+\mbox{ .1500E-01}}_{\,-\mbox{ .1500E-01}}$}
 & Ben66 \\
 {$\mbox{ .6000E+01}$}
 & {${\mbox{ .8920E-03}}^{\,+\mbox{ .1151E-03}}_{\,-\mbox{ .1151E-03}}$}
 & {${\mbox{ .1990E+00}}^{\,+\mbox{ .1283E-01}}_{\,-\mbox{ .1283E-01}}$}
 & Buc65 \\
 {$\mbox{ .6720E+01}$}
 & {${\mbox{ .8510E-03}}^{\,+\mbox{ .7914E-04}}_{\,-\mbox{ .7914E-04}}$}
 & {${\mbox{ .1834E+00}}^{\,+\mbox{ .8530E-02}}_{\,-\mbox{ .8530E-02}}$}
 & Auf85 \\
 {$\mbox{ .7000E+01}$}
 & {${\mbox{ .7390E-03}}^{\,+\mbox{ .9090E-04}}_{\,-\mbox{ .9090E-04}}$}
 & {${\mbox{ .1674E+00}}^{\,+\mbox{ .1030E-01}}_{\,-\mbox{ .1030E-01}}$}
 & Buc65 \\
 {$\mbox{ .7940E+01}$}
 & {${\mbox{ .5660E-03}}^{\,+\mbox{ .4811E-04}}_{\,-\mbox{ .4811E-04}}$}
 & {${\mbox{ .1374E+00}}^{\,+\mbox{ .5840E-02}}_{\,-\mbox{ .5840E-02}}$}
 & Auf85 \\
 {$\mbox{ .8000E+01}$}
 & {${\mbox{ .5230E-03}}^{\,+\mbox{ .6119E-04}}_{\,-\mbox{ .6119E-04}}$}
 & {${\mbox{ .1316E+00}}^{\,+\mbox{ .7697E-02}}_{\,-\mbox{ .7697E-02}}$}
 & Buc65 \\
 {$\mbox{ .9250E+01}$}
 & {${\mbox{ .4170E-03}}^{\,+\mbox{ .3378E-04}}_{\,-\mbox{ .3378E-04}}$}
 & {${\mbox{ .1091E+00}}^{\,+\mbox{ .4417E-02}}_{\,-\mbox{ .4417E-02}}$}
 & Auf85 \\
 {$\mbox{ .1086E+02}$}
 & {${\mbox{ .2520E-03}}^{\,+\mbox{ .2268E-04}}_{\,-\mbox{ .2268E-04}}$}
 & {${\mbox{ .7808E-01}}^{\,+\mbox{ .3514E-02}}_{\,-\mbox{ .3514E-02}}$}
 & Auf85 \\
 {$\mbox{ .1200E+02}$}
 & {${\mbox{ .1640E-03}}^{\,+\mbox{ .3198E-04}}_{\,-\mbox{ .3198E-04}}$}
 & {${\mbox{ .5983E-01}}^{\,+\mbox{ .5834E-02}}_{\,-\mbox{ .5834E-02}}$}
 & Buc65 \\
 {$\mbox{ .1255E+02}$}
 & {${\mbox{ .1720E-03}}^{\,+\mbox{ .8084E-04}}_{\,-\mbox{ .8084E-04}}$}
 & {${\mbox{ .5987E-01}}^{\,+\mbox{ .1407E-01}}_{\,-\mbox{ .1407E-01}}$}
 & Auf85 \\
 {$\mbox{ .1284E+02}$}
 & {${\mbox{ .1250E-03}}^{\,+\mbox{ .2900E-04}}_{\,-\mbox{ .2900E-04}}$}
 & {${\mbox{ .5054E-01}}^{\,+\mbox{ .5993E-02}}_{\,-\mbox{ .5993E-02}}$}
 & Cra85 \\
 {$\mbox{ .1459E+02}$}
 & {${\mbox{ .9380E-04}}^{\,+\mbox{ .9099E-05}}_{\,-\mbox{ .9099E-05}}$}
 & {${\mbox{ .4090E-01}}^{\,+\mbox{ .1983E-02}}_{\,-\mbox{ .1983E-02}}$}
 & Auf85 \\
 {$\mbox{ .1541E+02}$}
 & {${\mbox{ .8090E-04}}^{\,+\mbox{ .1470E-04}}_{\,-\mbox{ .1470E-04}}$}
 & {${\mbox{ .3696E-01}}^{\,+\mbox{ .3396E-02}}_{\,-\mbox{ .3396E-02}}$}
 & Cra85 \\
 {$\mbox{ .1618E+02}$}
 & {${\mbox{ .6460E-04}}^{\,+\mbox{ .5814E-05}}_{\,-\mbox{ .5814E-05}}$}
 & {${\mbox{ .3216E-01}}^{\,+\mbox{ .1447E-02}}_{\,-\mbox{ .1447E-02}}$}
 & Auf85 \\
 {$\mbox{ .1810E+02}$}
 & {${\mbox{ .3340E-04}}^{\,+\mbox{ .3641E-05}}_{\,-\mbox{ .3641E-05}}$}
 & {${\mbox{ .2181E-01}}^{\,+\mbox{ .1189E-02}}_{\,-\mbox{ .1189E-02}}$}
 & Auf85 \\
 {$\mbox{ .2003E+02}$}
 & {${\mbox{ .2810E-04}}^{\,+\mbox{ .1670E-04}}_{\,-\mbox{ .1670E-04}}$}
 & {${\mbox{ .1896E-01}}^{\,+\mbox{ .4155E-02}}_{\,-\mbox{ .4155E-02}}$}
 & Cra85 \\
 {$\mbox{ .2009E+02}$}
 & {${\mbox{ .2350E-04}}^{\,+\mbox{ .2491E-05}}_{\,-\mbox{ .2491E-05}}$}
 & {${\mbox{ .1732E-01}}^{\,+\mbox{ .9179E-03}}_{\,-\mbox{ .9179E-03}}$}
 & Auf85 \\
 {$\mbox{ .2184E+02}$}
 & {${\mbox{ .1510E-04}}^{\,+\mbox{ .1978E-05}}_{\,-\mbox{ .1978E-05}}$}
 & {${\mbox{ .1328E-01}}^{\,+\mbox{ .8701E-03}}_{\,-\mbox{ .8701E-03}}$}
 & Auf85 \\
 {$\mbox{ .2394E+02}$}
 & {${\mbox{ .7940E-05}}^{\,+\mbox{ .1286E-05}}_{\,-\mbox{ .1286E-05}}$}
 & {${\mbox{ .9175E-02}}^{\,+\mbox{ .7432E-03}}_{\,-\mbox{ .7432E-03}}$}
 & Auf85 \\
 {$\mbox{ .2568E+02}$}
 & {${\mbox{ .9480E-05}}^{\,+\mbox{ .3000E-05}}_{\,-\mbox{ .3000E-05}}$}
 & {${\mbox{ .9649E-02}}^{\,+\mbox{ .1538E-02}}_{\,-\mbox{ .1538E-02}}$}
 & Cra85 \\
 {$\mbox{ .2609E+02}$}
 & {${\mbox{ .6330E-05}}^{\,+\mbox{ .8989E-06}}_{\,-\mbox{ .8989E-06}}$}
 & {${\mbox{ .7826E-02}}^{\,+\mbox{ .5556E-03}}_{\,-\mbox{ .5556E-03}}$}
 & Auf85 \\
 {$\mbox{ .2797E+02}$}
 & {${\mbox{ .3330E-05}}^{\,+\mbox{ .8092E-06}}_{\,-\mbox{ .8092E-06}}$}
 & {${\mbox{ .5469E-02}}^{\,+\mbox{ .6644E-03}}_{\,-\mbox{ .6644E-03}}$}
 & Auf85 \\
 {$\mbox{ .3108E+02}$}
 & {${\mbox{ .1260E-05}}^{\,+\mbox{ .2400E-06}}_{\,-\mbox{ .2400E-06}}$}
 & {${\mbox{ .3179E-02}}^{\,+\mbox{ .3027E-03}}_{\,-\mbox{ .3027E-03}}$}
 & Arn87 \\
 {$\mbox{ .3339E+02}$}
 & {${\mbox{ .1870E-05}}^{\,+\mbox{ .9300E-06}}_{\,-\mbox{ .9300E-06}}$}
 & {${\mbox{ .3716E-02}}^{\,+\mbox{ .9189E-03}}_{\,-\mbox{ .9189E-03}}$}
 & Cra85 \\
 {$\mbox{ .3827E+02}$}
 & {${\mbox{ .3600E-06}}^{\,+\mbox{ .7000E-07}}_{\,-\mbox{ .7000E-07}}$}
 & {${\mbox{ .1517E-02}}^{\,+\mbox{ .1475E-03}}_{\,-\mbox{ .1475E-03}}$}
 & Arn87 \\
 {$\mbox{ .4135E+02}$}
 & {${\mbox{ .1220E-06}}^{\,+\mbox{ .3600E-07}}_{\,-\mbox{ .3600E-07}}$}
 & {${\mbox{ .8465E-03}}^{\,+\mbox{ .1249E-03}}_{\,-\mbox{ .1249E-03}}$}
 & Arn87 \\
 {$\mbox{ .4469E+02}$}
 & {${\mbox{ .1700E-07}}^{\,+\mbox{ .1500E-07}}_{\,-\mbox{ .1500E-07}}$}
 & {${\mbox{ .3027E-03}}^{\,+\mbox{ .1335E-03}}_{\,-\mbox{ .1335E-03}}$}
 & Arn87 \\
 {$\mbox{ .5085E+02}$}
 & {${\mbox{ .8000E-08}}^{\,+\mbox{ .9000E-08}}_{\,-\mbox{ .9000E-08}}$}
 & {${\mbox{ .1932E-03}}^{\,+\mbox{ .1087E-03}}_{\,-\mbox{ .1087E-03}}$}
 & Arn87 \\
 {$\mbox{ .5727E+02}$}
 & {${\mbox{ .1100E-07}}^{\,+\mbox{ .6000E-08}}_{\,-\mbox{ .6000E-08}}$}
 & {${\mbox{ .2118E-03}}^{\,+\mbox{ .5777E-04}}_{\,-\mbox{ .5777E-04}}$}
 & Arn87 \\
 {$\mbox{ .6369E+02}$}
 & {${\mbox{ .1500E-07}}^{\,+\mbox{ .8000E-08}}_{\,-\mbox{ .8000E-08}}$}
 & {${\mbox{ .2328E-03}}^{\,+\mbox{ .6207E-04}}_{\,-\mbox{ .6207E-04}}$}
 & Arn87 \\
 {$\mbox{ .6498E+02}$}
 & {${\mbox{ .1900E-07}}^{\,+\mbox{ .1100E-07}}_{\,-\mbox{ .1100E-07}}$}
 & {${\mbox{ .2590E-03}}^{\,+\mbox{ .7497E-04}}_{\,-\mbox{ .7497E-04}}$}
 & Arn87 \\
 {$\mbox{ .7114E+02}$}
 & {${\mbox{ .3000E-08}}^{\,+\mbox{ .5000E-08}}_{\,-\mbox{ .5000E-08}}$}
 & {${\mbox{ .9765E-04}}^{\,+\mbox{ .8137E-04}}_{\,-\mbox{ .8137E-04}}$}
 & Arn87 \\
\end{tabular}

\begin{tabular}{rccl}
 & & & \\
 {$k_z^2$ [${\mbox{fm}}^{-2}$] }
 & {$F_C(k_z^2)$}
 & {$F_Q(k_z^2)$}
 & Ref. \\ \hline \hline
 & & & \\
 {$\mbox{ .9761E+00}$}
 & {${\mbox{ .5500E+00}}^{\,+\mbox{ .7000E-02}}_{\,-\mbox{ .7000E-02}}$}
 & {${\mbox{ .3140E+02}}^{\,+\mbox{ .1440E+02}}_{\,-\mbox{ .1440E+02}}$}
 & Dmi85 \\
 {$\mbox{ .1831E+01}$}
 & {${\mbox{ .3770E+00}}^{\,+\mbox{ .3000E-02}}_{\,-\mbox{ .3000E-02}}$}
 & {${\mbox{ .6850E+01}}^{\,+\mbox{ .2680E+01}}_{\,-\mbox{ .2680E+01}}$}
 & Dmi85 \\
 {$\mbox{ .3000E+01}$}
 & {${\mbox{ .2550E+00}}^{\,+\mbox{ .7000E-02}}_{\,-\mbox{ .7000E-02}}$}
 & {${\mbox{ .7019E+01}}^{\,+\mbox{ .2322E+00}}_{\,-\mbox{ .2322E+00}}$}
 & Ben66 \\
 {$\mbox{ .3028E+01}$}
 & {${\mbox{ .2420E+00}}^{\,+\mbox{ .2000E-02}}_{\,-\mbox{ .2000E-02}}$}
 & {${\mbox{ .6130E+01}}^{\,+\mbox{ .9000E+00}}_{\,-\mbox{ .9000E+00}}$}
 & Sch84 \\
 {$\mbox{ .4000E+01}$}
 & {${\mbox{ .1710E+00}}^{\,+\mbox{ .5000E-02}}_{\,-\mbox{ .5000E-02}}$}
 & {${\mbox{ .4851E+01}}^{\,+\mbox{ .1548E+00}}_{\,-\mbox{ .1548E+00}}$}
 & Ben66 \\
 {$\mbox{ .4121E+01}$}
 & {${\mbox{ .1630E+00}}^{\,+\mbox{ .3000E-02}}_{\,-\mbox{ .4000E-02}}$}
 & {${\mbox{ .4430E+01}}^{\,+\mbox{ .1020E+01}}_{\,-\mbox{ .1020E+01}}$}
 & Sch84 \\
 {$\mbox{ .5000E+01}$}
 & {${\mbox{ .1250E+00}}^{\,+\mbox{ .3500E-02}}_{\,-\mbox{ .3500E-02}}$}
 & {${\mbox{ .3664E+01}}^{\,+\mbox{ .1032E+00}}_{\,-\mbox{ .1032E+00}}$}
 & Ben66 \\
 {$\mbox{ .6000E+01}$}
 & {${\mbox{ .9300E-01}}^{\,+\mbox{ .8000E-02}}_{\,-\mbox{ .8000E-02}}$}
 & {${\mbox{ .2838E+01}}^{\,+\mbox{ .2322E+00}}_{\,-\mbox{ .2322E+00}}$}
 & Ben66 \\
 {$\mbox{ .6200E+01}$}
 & {${\mbox{ .8670E-01}}^{\,+\mbox{ .2900E-02}}_{\,-\mbox{ .3900E-02}}$}
 & {${\mbox{ .2160E+01}}^{\,+\mbox{ .5000E+00}}_{\,-\mbox{ .5000E+00}}$}
 & Gil90 \\
 {$\mbox{ .8585E+01}$}
 & {${\mbox{ .3400E-01}}^{\,+\mbox{ .1270E-01}}_{\,-\mbox{ .2510E-01}}$}
 & {${\mbox{ .1840E+01}}^{\,+\mbox{ .5000E+00}}_{\,-\mbox{ .6500E+00}}$}
 & Gil90 \\
 {$\mbox{ .1429E+02}$}
 & {${\mbox{ .1270E-01}}^{\,+\mbox{ .4700E-02}}_{\,-\mbox{ .5600E-02}}$}
 & {${\mbox{ .4820E+00}}^{\,+\mbox{ .7700E-01}}_{\,-\mbox{ .1160E+00}}$}
 & The91 \\
 {$\mbox{ .1429E+02}$}
 & {${\mbox{ .1270E-01}}^{\,+\mbox{ .4700E-02}}_{\,-\mbox{ .5600E-02}}$}
 & {${\mbox{ .4820E+00}}^{\,+\mbox{ .7700E-01}}_{\,-\mbox{ .1160E+00}}$}
 & Gar94 \\
 {$\mbox{ .1781E+02}$}
 & {${\mbox{ .1660E-02}}^{\,+\mbox{ .1610E-02}}_{\,-\mbox{ .1420E-02}}$}
 & {${\mbox{ .3150E+00}}^{\,+\mbox{ .1000E-01}}_{\,-\mbox{ .1100E-01}}$}
 & The91 \\
 {$\mbox{ .1781E+02}$}
 & {${\mbox{ .1660E-02}}^{\,+\mbox{ .1610E-02}}_{\,-\mbox{ .1420E-02}}$}
 & {${\mbox{ .3150E+00}}^{\,+\mbox{ .1000E-01}}_{\,-\mbox{ .1100E-01}}$}
 & Gar94 \\
 {$\mbox{ .2134E+02}$}
 & {${\mbox{-.1470E-02}}^{\,+\mbox{ .1060E-02}}_{\,-\mbox{ .1040E-02}}$}
 & {${\mbox{ .1890E+00}}^{\,+\mbox{ .7000E-02}}_{\,-\mbox{ .8000E-02}}$}
 & The91 \\
 {$\mbox{ .2134E+02}$}
 & {${\mbox{-.1470E-02}}^{\,+\mbox{ .1060E-02}}_{\,-\mbox{ .1040E-02}}$}
 & {${\mbox{ .1890E+00}}^{\,+\mbox{ .7000E-02}}_{\,-\mbox{ .8000E-02}}$}
 & Gar94 \\
\end{tabular}

\begin{tabular}{rccl}
 & & & \\
 {$k_z$ [${\mbox{fm}}^{-1}$] }
 & {$\tilde{t}_{\,20}$}
 & {$t_{\,20} \,(70^{\circ})$}
 & Ref. \\ \hline \hline
 & & & \\
 {$\mbox{ .9761E+00}$}
 & {${\mbox{-.3000E+00}}^{\,+\mbox{ .1400E+00}}_{\,-\mbox{ .1400E+00}}$}
 & {${\mbox{-.3000E+00}}^{\,+\mbox{ .1400E+00}}_{\,-\mbox{ .1400E+00}}$}
 & Dmi85 \\
 {$\mbox{ .1831E+01}$}
 & {${\mbox{-.1800E+00}}^{\,+\mbox{ .7000E-01}}_{\,-\mbox{ .7000E-01}}$}
 & {${\mbox{-.1800E+00}}^{\,+\mbox{ .7000E-01}}_{\,-\mbox{ .7000E-01}}$}
 & Dmi85 \\
 {$\mbox{ .3028E+01}$}
 & {${\mbox{-.4100E+00}}^{\,+\mbox{ .6000E-01}}_{\,-\mbox{ .6000E-01}}$}
 & {${\mbox{-.4100E+00}}^{\,+\mbox{ .6000E-01}}_{\,-\mbox{ .6000E-01}}$}
 & Sch84 \\
 {$\mbox{ .4121E+01}$}
 & {${\mbox{-.5900E+00}}^{\,+\mbox{ .1400E+00}}_{\,-\mbox{ .1400E+00}}$}
 & {${\mbox{-.5800E+00}}^{\,+\mbox{ .1300E+00}}_{\,-\mbox{ .1300E+00}}$}
 & Sch84 \\
 {$\mbox{ .6200E+01}$}
 & {${\mbox{-.7900E+00}}^{\,+\mbox{ .1800E+00}}_{\,-\mbox{ .1800E+00}}$}
 & {${\mbox{-.7500E+00}}^{\,+\mbox{ .1600E+00}}_{\,-\mbox{ .1600E+00}}$}
 & Gil90 \\
 {$\mbox{ .8585E+01}$}
 & {${\mbox{-.1400E+01}}^{\,+\mbox{ .3700E+00}}_{\,-\mbox{ .3700E+00}}$}
 & {${\mbox{-.1260E+01}}^{\,+\mbox{ .3200E+00}}_{\,-\mbox{ .3200E+00}}$}
 & Gil90 \\
 {$\mbox{ .1429E+02}$}
 & {${\mbox{-.1450E+01}}^{\,+\mbox{ .1800E+00}}_{\,-\mbox{ .1800E+00}}$}
 & {${\mbox{-.1230E+01}}^{\,+\mbox{ .1800E+00}}_{\,-\mbox{ .1800E+00}}$}
 & Gar94 \\
 {$\mbox{ .1781E+02}$}
 & {${\mbox{-.9200E+00}}^{\,+\mbox{ .1800E+00}}_{\,-\mbox{ .1800E+00}}$}
 & {${\mbox{-.8300E+00}}^{\,+\mbox{ .1600E+00}}_{\,-\mbox{ .1600E+00}}$}
 & Gar94 \\
 {$\mbox{ .2134E+02}$}
 & {${\mbox{-.4200E+00}}^{\,+\mbox{ .2100E+00}}_{\,-\mbox{ .2100E+00}}$}
 & {${\mbox{-.4100E+00}}^{\,+\mbox{ .1800E+00}}_{\,-\mbox{ .1800E+00}}$}
 & Gar94 \\
\end{tabular}

\setlength{\columnsep}{3mm}

\twocolumn

\begin{tabular}{rcl}
 {$k_z^2$ [${\mbox{fm}}^{-2}$] }
 & {$A(k_z^2)$}
 & Ref. \\ \hline \hline
 & & \\
 {$\mbox{.4400E-01}$}
 & {${\mbox{.9358E+00}}^{\,+\mbox{.1500E-02}}_{\,-\mbox{.1500E-02}}$}
 & Sim81 \\
 {$\mbox{.2110E+00}$}
 & {${\mbox{.7450E+00}}^{\,+\mbox{.2800E-02}}_{\,-\mbox{.2800E-02}}$}
 & Sim81 \\
 {$\mbox{.2500E+00}$}
 & {${\mbox{.7101E+00}}^{\,+\mbox{.1500E-02}}_{\,-\mbox{.1500E-02}}$}
 & Sim81 \\
 {$\mbox{.3000E+00}$}
 & {${\mbox{.6659E+00}}^{\,+\mbox{.1600E-02}}_{\,-\mbox{.1600E-02}}$}
 & Sim81 \\
 {$\mbox{.3500E+00}$}
 & {${\mbox{.6220E+00}}^{\,+\mbox{.1600E-02}}_{\,-\mbox{.1600E-02}}$}
 & Sim81 \\
 {$\mbox{.4000E+00}$}
 & {${\mbox{.5911E+00}}^{\,+\mbox{.1500E-02}}_{\,-\mbox{.1500E-02}}$}
 & Sim81 \\
 {$\mbox{.4500E+00}$}
 & {${\mbox{.5548E+00}}^{\,+\mbox{.1400E-02}}_{\,-\mbox{.1400E-02}}$}
 & Sim81 \\
 {$\mbox{.4600E+00}$}
 & {${\mbox{.5455E+00}}^{\,+\mbox{.9300E-02}}_{\,-\mbox{.9300E-02}}$}
 & Pla90 \\
 {$\mbox{.5000E+00}$}
 & {${\mbox{.5247E+00}}^{\,+\mbox{.2000E-02}}_{\,-\mbox{.2000E-02}}$}
 & Sim81 \\
 {$\mbox{.5500E+00}$}
 & {${\mbox{.4972E+00}}^{\,+\mbox{.2200E-02}}_{\,-\mbox{.2200E-02}}$}
 & Sim81 \\
 {$\mbox{.6000E+00}$}
 & {${\mbox{.4662E+00}}^{\,+\mbox{.2800E-02}}_{\,-\mbox{.2800E-02}}$}
 & Sim81 \\
 {$\mbox{.6000E+00}$}
 & {${\mbox{.4698E+00}}^{\,+\mbox{.7100E-02}}_{\,-\mbox{.7100E-02}}$}
 & Pla90 \\
 {$\mbox{.7000E+00}$}
 & {${\mbox{.4198E+00}}^{\,+\mbox{.1800E-02}}_{\,-\mbox{.1800E-02}}$}
 & Sim81 \\
 {$\mbox{.7000E+00}$}
 & {${\mbox{.4148E+00}}^{\,+\mbox{.7500E-02}}_{\,-\mbox{.7500E-02}}$}
 & Pla90 \\
 {$\mbox{.8100E+00}$}
 & {${\mbox{.3743E+00}}^{\,+\mbox{.6400E-02}}_{\,-\mbox{.6400E-02}}$}
 & Pla90 \\
 {$\mbox{.9700E+00}$}
 & {${\mbox{.3107E+00}}^{\,+\mbox{.5600E-02}}_{\,-\mbox{.5600E-02}}$}
 & Pla90 \\
 {$\mbox{.1000E+01}$}
 & {${\mbox{.3103E+00}}^{\,+\mbox{.1300E-02}}_{\,-\mbox{.1300E-02}}$}
 & Sim81 \\
 {$\mbox{.1040E+01}$}
 & {${\mbox{.2954E+00}}^{\,+\mbox{.4400E-02}}_{\,-\mbox{.4400E-02}}$}
 & Pla90 \\
 {$\mbox{.1250E+01}$}
 & {${\mbox{.2387E+00}}^{\,+\mbox{.4500E-02}}_{\,-\mbox{.4500E-02}}$}
 & Pla90 \\
 {$\mbox{.1550E+01}$}
 & {${\mbox{.1911E+00}}^{\,+\mbox{.1100E-02}}_{\,-\mbox{.1100E-02}}$}
 & Sim81 \\
 {$\mbox{.1550E+01}$}
 & {${\mbox{.1778E+00}}^{\,+\mbox{.3400E-02}}_{\,-\mbox{.3400E-02}}$}
 & Pla90 \\
 {$\mbox{.1550E+01}$}
 & {${\mbox{.1858E+00}}^{\,+\mbox{.2800E-02}}_{\,-\mbox{.2800E-02}}$}
 & Pla90 \\
 {$\mbox{.1840E+01}$}
 & {${\mbox{.1400E+00}}^{\,+\mbox{.2400E-02}}_{\,-\mbox{.2400E-02}}$}
 & Pla90 \\
 {$\mbox{.2100E+01}$}
 & {${\mbox{.1241E+00}}^{\,+\mbox{.7000E-03}}_{\,-\mbox{.7000E-03}}$}
 & Sim81 \\
 {$\mbox{.2130E+01}$}
 & {${\mbox{.1127E+00}}^{\,+\mbox{.2200E-02}}_{\,-\mbox{.2200E-02}}$}
 & Pla90 \\
 {$\mbox{.2130E+01}$}
 & {${\mbox{.1140E+00}}^{\,+\mbox{.1700E-02}}_{\,-\mbox{.1700E-02}}$}
 & Pla90 \\
 {$\mbox{.2210E+01}$}
 & {${\mbox{.1109E+00}}^{\,+\mbox{.1700E-02}}_{\,-\mbox{.1700E-02}}$}
 & Pla90 \\
 {$\mbox{.2390E+01}$}
 & {${\mbox{.9145E-01}}^{\,+\mbox{.2010E-02}}_{\,-\mbox{.2010E-02}}$}
 & Pla90 \\
 {$\mbox{.2630E+01}$}
 & {${\mbox{.7825E-01}}^{\,+\mbox{.2040E-02}}_{\,-\mbox{.2040E-02}}$}
 & Pla90 \\
 {$\mbox{.2740E+01}$}
 & {${\mbox{.7249E-01}}^{\,+\mbox{.1200E-03}}_{\,-\mbox{.1200E-03}}$}
 & Pla90 \\
 {$\mbox{.2820E+01}$}
 & {${\mbox{.7014E-01}}^{\,+\mbox{.1050E-02}}_{\,-\mbox{.1050E-02}}$}
 & Pla90 \\
 {$\mbox{.2850E+01}$}
 & {${\mbox{.6633E-01}}^{\,+\mbox{.1920E-02}}_{\,-\mbox{.1920E-02}}$}
 & Pla90 \\
 \end{tabular}

\pagebreak

\begin{tabular}{rcl}
 {$k_z^2$ [${\mbox{fm}}^{-2}$] }
 & {$A(k_z^2)$}
 & Ref. \\ \hline \hline
 & & \\
 {$\mbox{.3300E+01}$}
 & {${\mbox{ .5390E-01}}^{\,+\mbox{.4000E-03}}_{\,-\mbox{.4000E-03}}$}
 & Sim81 \\
 {$\mbox{.3360E+01}$}
 & {${\mbox{ .4844E-01}}^{\,+\mbox{.8700E-03}}_{\,-\mbox{.8700E-03}}$}
 & Pla90 \\
 {$\mbox{.3480E+01}$}
 & {${\mbox{ .4478E-01}}^{\,+\mbox{.6700E-03}}_{\,-\mbox{.6700E-03}}$}
 & Pla90 \\
 {$\mbox{.3680E+01}$}
 & {${\mbox{ .3928E-01}}^{\,+\mbox{.6700E-03}}_{\,-\mbox{.6700E-03}}$}
 & Pla90 \\
 {$\mbox{.3960E+01}$}
 & {${\mbox{ .3233E-01}}^{\,+\mbox{.6100E-03}}_{\,-\mbox{.6100E-03}}$}
 & Pla90 \\
 {$\mbox{.4000E+01}$}
 & {${\mbox{ .3550E-01}}^{\,+\mbox{.4000E-03}}_{\,-\mbox{.4000E-03}}$}
 & Sim81 \\
 {$\mbox{.4170E+01}$}
 & {${\mbox{ .2874E-01}}^{\,+\mbox{.4300E-03}}_{\,-\mbox{.4300E-03}}$}
 & Pla90 \\
 {$\mbox{.4540E+01}$}
 & {${\mbox{ .2282E-01}}^{\,+\mbox{.5000E-03}}_{\,-\mbox{.5000E-03}}$}
 & Pla90 \\
 {$\mbox{.4680E+01}$}
 & {${\mbox{ .2072E-01}}^{\,+\mbox{.3100E-03}}_{\,-\mbox{.3100E-03}}$}
 & Pla90 \\
 {$\mbox{.4900E+01}$}
 & {${\mbox{ .1856E-01}}^{\,+\mbox{.2800E-03}}_{\,-\mbox{.2800E-03}}$}
 & Pla90 \\
 {$\mbox{.5650E+01}$}
 & {${\mbox{ .1247E-01}}^{\,+\mbox{.1900E-03}}_{\,-\mbox{.1900E-03}}$}
 & Pla90 \\
 {$\mbox{.5760E+01}$}
 & {${\mbox{ .1164E-01}}^{\,+\mbox{.1800E-03}}_{\,-\mbox{.1800E-03}}$}
 & Pla90 \\
 {$\mbox{.6000E+01}$}
 & {${\mbox{ .9180E-02}}^{\,+\mbox{.5324E-03}}_{\,-\mbox{.5324E-03}}$}
 & Buc65 \\
 {$\mbox{.6160E+01}$}
 & {${\mbox{ .9860E-02}}^{\,+\mbox{.5000E-03}}_{\,-\mbox{.5000E-03}}$}
 & Gal71 \\
 {$\mbox{.6310E+01}$}
 & {${\mbox{ .8904E-02}}^{\,+\mbox{.1690E-03}}_{\,-\mbox{.1690E-03}}$}
 & Pla90 \\
 {$\mbox{.6400E+01}$}
 & {${\mbox{ .8421E-02}}^{\,+\mbox{.1600E-03}}_{\,-\mbox{.1600E-03}}$}
 & Pla90 \\
 {$\mbox{.6600E+01}$}
 & {${\mbox{ .7570E-02}}^{\,+\mbox{.4500E-03}}_{\,-\mbox{.4500E-03}}$}
 & Gal71 \\
 {$\mbox{.6880E+01}$}
 & {${\mbox{ .6730E-02}}^{\,+\mbox{.1010E-03}}_{\,-\mbox{.1010E-03}}$}
 & Pla90 \\
 {$\mbox{.7000E+01}$}
 & {${\mbox{ .5710E-02}}^{\,+\mbox{.3198E-03}}_{\,-\mbox{.3198E-03}}$}
 & Buc65 \\
 {$\mbox{.7040E+01}$}
 & {${\mbox{ .6400E-02}}^{\,+\mbox{.4100E-03}}_{\,-\mbox{.4100E-03}}$}
 & Gal71 \\
 {$\mbox{.7160E+01}$}
 & {${\mbox{ .5743E-02}}^{\,+\mbox{.9300E-04}}_{\,-\mbox{.9300E-04}}$}
 & Pla90 \\
 {$\mbox{.7460E+01}$}
 & {${\mbox{ .5028E-02}}^{\,+\mbox{.7500E-04}}_{\,-\mbox{.7500E-04}}$}
 & Pla90 \\
 {$\mbox{.7500E+01}$}
 & {${\mbox{ .5390E-02}}^{\,+\mbox{.3800E-03}}_{\,-\mbox{.3800E-03}}$}
 & Gal71 \\
 {$\mbox{.7920E+01}$}
 & {${\mbox{ .4169E-02}}^{\,+\mbox{.6700E-04}}_{\,-\mbox{.6700E-04}}$}
 & Pla90 \\
 {$\mbox{.8000E+01}$}
 & {${\mbox{ .3630E-02}}^{\,+\mbox{.1815E-03}}_{\,-\mbox{.1815E-03}}$}
 & Buc65 \\
 {$\mbox{.8040E+01}$}
 & {${\mbox{ .3958E-02}}^{\,+\mbox{.5900E-04}}_{\,-\mbox{.5900E-04}}$}
 & Pla90 \\
 {$\mbox{.8670E+01}$}
 & {${\mbox{ .3151E-02}}^{\,+\mbox{.8200E-04}}_{\,-\mbox{.8200E-04}}$}
 & Pla90 \\
 {$\mbox{.9220E+01}$}
 & {${\mbox{ .2413E-02}}^{\,+\mbox{.5100E-04}}_{\,-\mbox{.5100E-04}}$}
 & Pla90 \\
 {$\mbox{.9400E+01}$}
 & {${\mbox{ .2343E-02}}^{\,+\mbox{.7000E-04}}_{\,-\mbox{.7000E-04}}$}
 & Pla90 \\
 {$\mbox{.9750E+01}$}
 & {${\mbox{ .2050E-02}}^{\,+\mbox{.1500E-03}}_{\,-\mbox{.1500E-03}}$}
 & Gal71 \\
 {$\mbox{.1041E+02}$}
 & {${\mbox{ .1562E-02}}^{\,+\mbox{.3600E-04}}_{\,-\mbox{.3600E-04}}$}
 & Pla90 \\
 {$\mbox{.1090E+02}$}
 & {${\mbox{ .1490E-02}}^{\,+\mbox{.1200E-03}}_{\,-\mbox{.1200E-03}}$}
 & Gal71 \\
 \end{tabular}

\begin{tabular}{rcl}
 {$k_z^2$ [${\mbox{fm}}^{-2}$] }
 & {$A(k_z^2)$}
 & Ref. \\ \hline \hline
 & & \\
 {$\mbox{.1130E+02}$}
 & {${\mbox{ .1287E-02}}^{\,+\mbox{.7800E-04}}_{\,-\mbox{.7800E-04}}$}
 & Gal71 \\
 {$\mbox{.1160E+02}$}
 & {${\mbox{ .1038E-02}}^{\,+\mbox{.2900E-04}}_{\,-\mbox{.2900E-04}}$}
 & Pla90 \\
 {$\mbox{.1170E+02}$}
 & {${\mbox{ .1077E-02}}^{\,+\mbox{.6400E-04}}_{\,-\mbox{.6400E-04}}$}
 & Gal71 \\
 {$\mbox{.1200E+02}$}
 & {${\mbox{ .8950E-03}}^{\,+\mbox{.1065E-03}}_{\,-\mbox{.1065E-03}}$}
 & Buc65 \\
 {$\mbox{.1242E+02}$}
 & {${\mbox{ .9260E-03}}^{\,+\mbox{.6800E-04}}_{\,-\mbox{.6800E-04}}$}
 & Gal71 \\
 {$\mbox{.1284E+02}$}
 & {${\mbox{ .7790E-03}}^{\,+\mbox{.3400E-04}}_{\,-\mbox{.3400E-04}}$}
 & Cra85 \\
 {$\mbox{.1300E+02}$}
 & {${\mbox{ .7790E-03}}^{\,+\mbox{.6300E-04}}_{\,-\mbox{.6300E-04}}$}
 & Gal71 \\
 {$\mbox{.1391E+02}$}
 & {${\mbox{ .5403E-03}}^{\,+\mbox{.1840E-04}}_{\,-\mbox{.1840E-04}}$}
 & Pla90 \\
 {$\mbox{.1442E+02}$}
 & {${\mbox{ .4560E-03}}^{\,+\mbox{.4150E-04}}_{\,-\mbox{.4150E-04}}$}
 & Eli69 \\
 {$\mbox{.1472E+02}$}
 & {${\mbox{ .4288E-03}}^{\,+\mbox{.3690E-04}}_{\,-\mbox{.3690E-04}}$}
 & Eli69 \\
 {$\mbox{.1541E+02}$}
 & {${\mbox{ .4630E-03}}^{\,+\mbox{.4500E-04}}_{\,-\mbox{.4500E-04}}$}
 & Cra85 \\
 {$\mbox{.1607E+02}$}
 & {${\mbox{ .3575E-03}}^{\,+\mbox{.3400E-04}}_{\,-\mbox{.3400E-04}}$}
 & Eli69 \\
 {$\mbox{.1608E+02}$}
 & {${\mbox{ .3499E-03}}^{\,+\mbox{.1570E-04}}_{\,-\mbox{.1570E-04}}$}
 & Pla90 \\
 {$\mbox{.1666E+02}$}
 & {${\mbox{ .2808E-03}}^{\,+\mbox{.2530E-04}}_{\,-\mbox{.2530E-04}}$}
 & Eli69 \\
 {$\mbox{.1729E+02}$}
 & {${\mbox{ .2760E-03}}^{\,+\mbox{.2580E-04}}_{\,-\mbox{.2580E-04}}$}
 & Eli69 \\
 {$\mbox{.1806E+02}$}
 & {${\mbox{ .2335E-03}}^{\,+\mbox{.1990E-04}}_{\,-\mbox{.1990E-04}}$}
 & Pla90 \\
 {$\mbox{.1807E+02}$}
 & {${\mbox{ .2166E-03}}^{\,+\mbox{.2110E-04}}_{\,-\mbox{.2110E-04}}$}
 & Eli69 \\
 {$\mbox{.1878E+02}$}
 & {${\mbox{ .1575E-03}}^{\,+\mbox{.1790E-04}}_{\,-\mbox{.1790E-04}}$}
 & Eli69 \\
 {$\mbox{.1952E+02}$}
 & {${\mbox{ .1926E-03}}^{\,+\mbox{.2060E-04}}_{\,-\mbox{.2060E-04}}$}
 & Eli69 \\
 {$\mbox{.2000E+02}$}
 & {${\mbox{ .1710E-03}}^{\,+\mbox{.4395E-04}}_{\,-\mbox{.4395E-04}}$}
 & Buc65 \\
 {$\mbox{.2003E+02}$}
 & {${\mbox{ .1770E-03}}^{\,+\mbox{.7000E-05}}_{\,-\mbox{.7000E-05}}$}
 & Cra85 \\
 {$\mbox{.2033E+02}$}
 & {${\mbox{ .1588E-03}}^{\,+\mbox{.1870E-04}}_{\,-\mbox{.1870E-04}}$}
 & Eli69 \\
 {$\mbox{.2105E+02}$}
 & {${\mbox{ .1302E-03}}^{\,+\mbox{.1480E-04}}_{\,-\mbox{.1480E-04}}$}
 & Eli69 \\
 {$\mbox{.2182E+02}$}
 & {${\mbox{ .1129E-03}}^{\,+\mbox{.1470E-04}}_{\,-\mbox{.1470E-04}}$}
 & Eli69 \\
 {$\mbox{.2258E+02}$}
 & {${\mbox{ .9330E-04}}^{\,+\mbox{.1500E-04}}_{\,-\mbox{.1500E-04}}$}
 & Eli69 \\
 {$\mbox{.2338E+02}$}
 & {${\mbox{ .1081E-03}}^{\,+\mbox{.1870E-04}}_{\,-\mbox{.1870E-04}}$}
 & Eli69 \\
 {$\mbox{.2425E+02}$}
 & {${\mbox{ .7010E-04}}^{\,+\mbox{.1760E-04}}_{\,-\mbox{.1760E-04}}$}
 & Eli69 \\
 {$\mbox{.2538E+02}$}
 & {${\mbox{ .6750E-04}}^{\,+\mbox{.1620E-04}}_{\,-\mbox{.1620E-04}}$}
 & Eli69 \\
 {$\mbox{.2568E+02}$}
 & {${\mbox{ .7100E-04}}^{\,+\mbox{.2100E-05}}_{\,-\mbox{.2100E-05}}$}
 & Cra85 \\
 {$\mbox{.2743E+02}$}
 & {${\mbox{ .4380E-04}}^{\,+\mbox{.1670E-04}}_{\,-\mbox{.1670E-04}}$}
 & Eli69 \\
 {$\mbox{.3090E+02}$}
 & {${\mbox{ .5400E-04}}^{\,+\mbox{.1490E-04}}_{\,-\mbox{.1490E-04}}$}
 & Eli69 \\
 {$\mbox{.3339E+02}$}
 & {${\mbox{ .1770E-04}}^{\,+\mbox{.2600E-05}}_{\,-\mbox{.2600E-05}}$}
 & Cra85 \\
 {$\mbox{.3410E+02}$}
 & {${\mbox{ .2570E-04}}^{\,+\mbox{.1340E-04}}_{\,-\mbox{.1340E-04}}$}
 & Eli69 \\
 \end{tabular}

\onecolumn

\end{appendix}
%
%
% Literaturverzeichnis:
% =====================
%

%

\end{document}